\documentclass[journal,twoside]{IEEEtrann}

\IEEEoverridecommandlockouts

\usepackage{amsmath,amssymb,amsfonts,amsthm}
\usepackage{textcomp}
\usepackage{xcolor}
\usepackage{graphicx}
\usepackage{cite}
\usepackage{color}
\usepackage{url}
\usepackage{balance}
\usepackage{soul}
\setstcolor{red}
\setulcolor{orange}

\usepackage{siunitx}
\newtheorem{assumption}{Assumption}
\usepackage{enumitem}

\newtheorem{Prop5}{Theorem}
\newtheorem{Prop4}{\emph{Corollary}}
\newtheorem{Remark}{Remark}

\usepackage{multicol}
\usepackage{multirow}
\usepackage{caption}
\usepackage{makecell}
\usepackage[subtle]{savetrees}
\usepackage{float}
\usepackage{stfloats}
\begin{document}

%%%%%%%%%%%%%%%%%%%%%%%%%%%%%%%%%%%%%%%
%%%%%%%%%%%%%%%%%%%%%%%%%%%%%%%%%%%%%%%

%\title{\vspace{-0.07in} DC Microgrids with Nested Nonlinear Distributed Control Frameworks: Scalable Large-Signal Stability Guarantees}

\title{\vspace{-0.07in} DC Microgrids with Nested Nonlinear Distributed Control:\\ Scalable Large-Signal Stability and Voltage Containment}

\author{Cornelia Skaga,~\IEEEmembership{Member,~IEEE}, Mahdieh S. Sadabadi ,~\IEEEmembership{Senior Member,~IEEE}, and Gilbert Bergna-Diaz,~\IEEEmembership{Member,~IEEE}
\vspace{-0.15in}

% Global exponential stability guratensees 
%Large-Signal Stability Guarantees for a \\Scalable DC Microgrid with a Nested Nonlinear Control Framework:\\ The Slow Communication Scenario 

%Distributed Current Sharing in DC Microgrids Under Voltage Limits With Scalable Stability Guarantees

%Fast com, 

%\thanks{This work was supported by the Department of Electric Energy, Norwegian University of Science and Technology (NTNU), under Grant\hl{ XXXXX. Paper no. XXXX-X-2025}. (Corresponding author: Cornelia Skaga.)}
\thanks{
Cornelia Skaga with the Department of Electric Energy, Norwegian University of Science and Technology (NTNU), 7491 Trondheim, Norway (e-mail: cornelia.skaga@ntnu.no). 

Mahdieh S. Sadabadi is with the Department of Electrical and Electronic
Engineering, The University of Manchester, M13 9PL, Manchester, UK (e-mail:
mahdieh.sadabadi@manchester.ac.uk).

Gilbert Bergna-Diaz is with the Department of Electric Energy, Norwegian University of Science and Technology (NTNU), 7491 Trondheim, Norway (e-mail: gilbert.bergna@ntnu.no).}
}
\markboth{Submitted for Publication}{Submitted for Publication}

\maketitle

%%%%%%%%%%%%%%%%%%%%%%%%%%%%%%%%%%%%%%%%%%%%%%%%%%%%%%%%%%%
%%%%%%%%%%%%%%%%%%%%%%%%%%%%%%%%%%%%%%%%%%%%%%%%%%%%%%%%%%%
\begin{abstract}
%As modern power systems continue to evolve into multi-agent, converter-dominated systems that demand scalable, stabilizing, and optimal control architectures,c} 
This paper investigates a cyber-physical DC microgrid employing a \emph{nonlinear} distributed consensus-based control scheme for coordinated integration and management of distributed generating units within an expandable framework. Relying on nested primary and secondary control loops--a (distributed) outer-loop and a (decentralized) inner-loop--the controller achieves proportional current sharing among all distributed generation units,
%\ul{realize} steady-state convergence where all energized units supply power proportionally to their rated capacities 
while dynamically operating within predefined \emph{voltage limits}. A rigorous Lyapunov-based stability analysis establishes a \emph{scalable} global exponential stability certificate under some tuning conditions and sufficient time-scale separation between the control loops--based on singular perturbation theory. An optimization-based tuning strategy is then formulated to identify and subsequently diminish unstable operating conditions. In turn, various practical tuning strategies are introduced to provide stable operations while facilitating \emph{near-optimal} proportional current sharing. The effectiveness of the proposed control framework and tuning approaches are finally supported through time-domain simulations of a case-specific low-voltage DC microgrid.
\end{abstract}

%Keeping the structure 
%The primary objective of future work is to develop a scalable stability proof that ensures stability without compromising the precise current sharing control objective.
%global exponential stability are ensured under appropriate tuning that verifies some \emph{stability conditions}, and the convenient mathematical structure of the proposed distributed control framework preserves scalable stability guarantees using Lyapunov arguments.}
%The converged equilibrium is then proven to satisfy optimal operations as both voltage containment and proportional current sharing are ensured within acceptable limits, 
%under an appropriate tuning that verifies some \emph{stability conditions}. It turns out that our proposal preserves a more convenient mathematical structure for the scalability of the stability proof using Lyapunov arguments. 

\begin{IEEEkeywords}
Nonlinear nested control, scalable global exponential stability, distributed cooperative action, voltage containment
\end{IEEEkeywords}
%%%%%%%%%%%%%%%%%%%%%%%%%%%%%%%%%%%%%%%%%%%%%%%%%%%%%%%%%%%
%%%%%%%%%%%%%%%%%%%%%%%%%%%%%%%%%%%%%%%%%%%%%%%%%%%%%%%%%%%

\section{Introduction}
Driven by escalating climate concerns and the imperative to decarbonize the energy sector, modern power systems are shifting from traditional large, fossil fuel-based generation to multiple smaller, distributed converter-interfaced renewable energy sources (RES) such as photovoltaics, wind turbines, and battery energy storage systems. This transition, however, introduces new control and stability challenges, motivating a paradigm shift from centralized to distributed control architectures \cite{ Poppi_J1, Poppi_J2, Av_Volt_3}. Distributed consensus-based control configurations leverage peer-to-peer communication among neighboring units as part of the secondary control layer, enabling coordinated decision-making and optimal system operation in accordance with the specified control objectives \cite{Av_Volt_3}. Driven by this paradigm shift, multi-agent microgrids (MGs)--small-scale electricity networks or distribution systems--have attracted significant research attention as an effective means to interconnect distributed generators (DGs), energy storage systems, and loads. These systems offer several operational advantages, including enhanced power flow management, reduced transmission losses, and the flexibility to operate in both grid-connected and islanded modes, while also enabling more active participation of consumers in energy markets \cite{Av_volt_1, Av_Volt_2, Detection_2}. When governed by distributed consensus-based control configurations, these MGs--often referred to as cyber-physical MGs (CPMGs) due to their reliance on communication infrastructures in the secondary controller--exhibit additional benefits such as improved scalability, enhanced reliability, and resilience against single points of failure \cite{Av_Volt_3}. Furthermore, considering the electrical characteristics of RES and modern generation units and loads,  DC CPMGs are increasingly recognized as a key enabler of the ongoing transition toward safe, stable, and sustainable electrification of the energy sector \cite{Babak_Set_Point}. From an industrial perspective, such DC MGs have proven beneficial in various applications, including the marine industry—such as shipboard microgrids and large ports—as well as in aircraft, railway systems, and large-scale charging facilities for electric vehicles\cite{Av_Volt_5, MArine_2, Marine_3, Marine_4}.

\subsection{Literature review and Research Gap}
Replacing traditional power systems—-dominated by AC synchronous generators with high system inertia-—with modern multi-agent systems that interconnect numerous smaller RES through power electronic converters (PEC)--characterized by low inertia and weak damping--increases the systems vulnerability to transient disturbances and reduces its robustness against random perturbations, thereby highlighting the imperative for efficient and stable control configurations \cite{modeling+stability_rev}. This transition also brings several operational challenges, particularly concerning voltage stability and power flow limitations \cite{SS_CS_Book_ch_6}. In the case of DC MGs, maintaining proper voltage regulation is crucial to ensure safe and reliable network operation, preventing voltage magnitudes from exceeding critical levels or experiencing sudden drops that could damage connected loads
%which typically depend on a supplier-defined nominal reference voltage 
\cite{Av_volt_1, Av_Volt_2}. Moreover, optimal power flow management--aimed at ensuring efficient power distribution across the network to satisfy load demands--requires proper load (power or current) sharing among DGs to avoid generator overloading, potential system failure, or even blackouts \cite{Av_Volt_2}. Given that multi-agent DC MGs interconnect DGs with diverse generation capacities, it is favorable that the total load demand is proportionally allocated among the active sources according to their respective capacities—an objective commonly achieved by regulating the voltage setpoints of the individual converters \cite{Av_Volt_5, Av_Volt_4}. %following a droop control characteristic 
Consequently, accurate voltage regulation and proportional load sharing (power- or current sharing) are typically considered the primary control objectives in DC MGs. However, due to the strong coupling between the voltage and current outputs of DGs, achieving proportional current sharing inherently constrains the ability to independently regulate the voltage at each node to its desired value. In other words, it is not possible to freely adjust node voltages while simultaneously maintaining proportional current sharing \cite{Av_Volt_4, Av_Volt_5}. 

To accommodate achieving both objectives simultaneously, hierarchical control structures are typically employed, generally organized into a three-level framework (primary, secondary and tertiary), allowing for a certain degree of independence between the control levels, operating at different time-scales and thereby varying control bandwidth requirements \cite{Hierarchical}.%\cite{Av_Volt_5, Av_Volt_2}
The PECs are primarily controlled by \emph{inner-loop} configurations--regulating the voltage or current/power output in accordance with a given reference value--and/or by \emph{droop control} characteristics.
%, which may provide a preliminary power-sharing mechanism. 
However, for the latter, a trade-off exists between power sharing performance and voltage regulation: an elevated droop gain enhances current-sharing performance while simultaneously amplifying the voltage deviation \cite{Hierarchical}. Hence, by relaxing the individual voltage requirements and instead focusing on maintaining an overall voltage level across the microgrid, \emph{voltage balancing} can be achieved through a primary control configuration—such as droop control—that limits deviations of the DG voltage outputs with respect to a global setpoint reference, typically defined as a weighted sum of the individual voltage references--also referred to as \emph{average voltage regulation}\cite{Av_Volt_5, Av_volt_1, Av_Volt_4}. Moreover, implementing a secondary control layer enables the adjustment of the voltage setpoints for the primary control loop, ensuring that the voltages of the DGs remain within a prescribed operating range to facilitate proportional current sharing \cite{Hierarchical}. Recent advancements have been directed toward distributed consensus-based secondary controllers, embedding digital communication links between neighboring units to enable coordinated control action among the active agents. This approach ensures that the \emph{optimal/desired} setpoint is delivered to the primary controller, facilitating proportional current sharing while simultaneously maintaining average voltage regulation at the primary control level \cite{Babak_Base_Article, Dist+droop_1, Dist+droop_2, Dist+droop_3, Dist+droop_4} or incorporated in the secondary control configurations\---see \cite{Babak_Set_Point, Av_Volt_5, Av_Volt_3, Av_volt_1, Av_Volt_4, Av_Volt_2, Av_6}, and references therein. 

Although average voltage regulation schemes operate satisfactory under nominal conditions, larger disturbances such as load and/or topological changes, may lead to significant deviations in individual generator voltages, potentially exceeding operational limits \cite{Babak_Set_Point}. To overcome these limitations, this research proposes a modified distributed \emph{nonlinear} consensus-based distributed control framework for DC MGs-- extending the idea initially proposed in \cite{Babak_Set_Point}, and further modified in \cite{Poppi_J1, Poppi_J2}. The proposed control framework relies on nested primary/secondary control loops-- a decentralized inner- and a distributed outer-loop--designed to dynamically contain the DGs output voltages within pre-specified thresholds while ensuring proportional current sharing in the steady state. Moreover, the distributed/decentralized structure of the control framework enables scalable applicability, even in the case of any communication link delay or failure. However, introducing the nonlinear characteristics in the inner-loop control architecture (to provide voltage containment of the individual PECs) complicates the search for a scalable global exponential stability (G.E.S.) certificate.

%However, the framework in \cite{Babak_Set_Point} was proposed without any stability guarantees. Furthermore, introducing the nonlinear characteristics in the inner-loop control architecture (to provide voltage containment of the individual PECs) complicates the search for a scalable global exponential stability (G.E.S.) certificate. 

\subsection{Contributions}
Our previous results \cite{Poppi_J1, Poppi_J2} have \emph{structurally modified} the initial contribution of \cite{Babak_Set_Point}---originally proposed without any stability proof--such that \emph{scalable} G.E.S. could be guaranteed. More precisely, we achieved this by
%\textcolor{gray}{In the previously proposed \emph{modified} versions \cite{Poppi_J1, Poppi_J2} of the nonlinear distributed control framework in \cite{Babak_Set_Point}, \emph{scalable} G.E.S was guaranteed by} 
first applying singular perturbation theory (SPT) under the assumption of sufficient time-scale separation between the electrical system including the decentralized inner-loop controllers and the distributed outer-loop controller-- motivated by the approach presented in \cite{Babak_AC}. Subsequently, Lyapunov theory is employed to construct separate Lyapunov functions for the two time-scaled systems, which are then combined into a composite Lyapunov function for the \emph{full} system. 

However, these frameworks guarantee stability only under relatively conservative conditions, where \cite{Poppi_J1} relies on significant timescale separation enforcing very fast communication rates, making the result more suitable for geographically contained systems. Additionally, the framework depends on an inner-loop leakage, that must be carefully tuned to adhere to the time-scale separation while ensuring that the \emph{optimal} proportional current sharing remains within an acceptable tolerance. Moreover, the stability certificate obtained in \cite{Poppi_J2}, relies on an inverted time-scale separation where the communications operates at a slower time-scale than the electrical system, combined with careful tuning of the \emph{non-permanent} leakage term and some specific electrical requirements. Hence, the results in \cite{Poppi_J1, Poppi_J2} are relatively conservative, application/system-dependent, and requires the communication dynamics to operate at a specific (slow/fast) operational rate. To deal with these disadvantages, in this work, we investigate an alternative approach by introducing a distinct time-scale separation that enforces time-scale requirements solely on the decentralized inner-loop controller. This eliminates the timescale-separation requirements between the communication and the electrical system, used in \cite{Poppi_J1, Poppi_J2}. Our goal is to determine whether our modified controller can reduce the conservatism of previous large stability margins and yield a global exponential stability certificate under less restrictive conditions. To facilitate this, we modify our control framework in the following ways:

%\ul{However, these frameworks guarantee stability only under relatively conservative conditions, relying on significant timescale separation and/or additional leakage and tuning constraints.} Thus, in this work, we investigate an alternative approach by introducing a distinct time-scale separation in which only the decentralized inner-loop operates at a slower timescale than the rest of the system. This relaxes the requirement that the outer-loop controller and its communication must be either faster or slower than the electrical system \st{and its decentralized controller}. Our goal is to determine whether our modified controller, under this time-scale separation, can reduce the conservatism of previous large stability margins and yield a global exponential stability certificate under less restrictive conditions. To facilitate this, we modify our control framework in the following ways:
\begin{itemize}[label=\textendash]
    \item We introduce a tunable leakage term in the control input of our distributed outer-loop controller to couple the electrical system with the communication dynamics--facilitating the construction of a quadratic Lyapunov function.
    \item We include a \emph{permanent} and individually tunable leakage in the decentralized inner-loop controller. 
    \item For mathematical feasibility we adjust the \emph{non-permanent} leakage in the decentralized controllers and include a leakage term in the distributed outer-loop.
    %\item We modify the \emph{non-permanent} leakage in the decentralized controllers by adjusting its slope and active contribution
    %\item For mathematical feasibility, a leakage term is also included in the distributed outer-loop; however, this term has a negligible influence on the power system dynamics.
\end{itemize}
Finally, for optimal performance, we propose three tuning strategies  with a different degrees of conservatism, scalability and performance. All tunings have been based on the Geršgorin circle theorem, aiming to guarantee global exponential stability, under voltage containment and achieving \emph{practical} proportional current sharing in steady state.

The rest of the paper is structured as follows. Section \ref{Model} presents the control objectives and proposed control framework, including the dynamical electrical DC MG model. In Section \ref{stability}, we present a comprehensive stability proof for a closed-loop DC microgrid with a nonlinear distributed controller, followed by Section \ref{sec_tuning_alg}, which proposes various tuning strategies to guarantee stability and convergence to a desired (optimal) steady state. Section \ref{sec_Case_Studies} further validates the derived stability conditions and proposed tuning strategies using a 4-terminal case-specific DC microgrid, followed by time-domain simulations that demonstrate the effectiveness of the proposed control framework.
%, followed by a small-signal stability analysis. 
Finally, Section \ref{conc} concludes this paper.  
%%%%%%%%%%%%%%%%%%%%%%%%%%%%%%%%%%%%%%%%%%%%%%%%%%%%%%%%%%%%%%
\section{System Modeling and Control Framework} \label{Model}
\begin{figure*}[!t]
    \centering
    \includegraphics[width=0.8\textwidth]{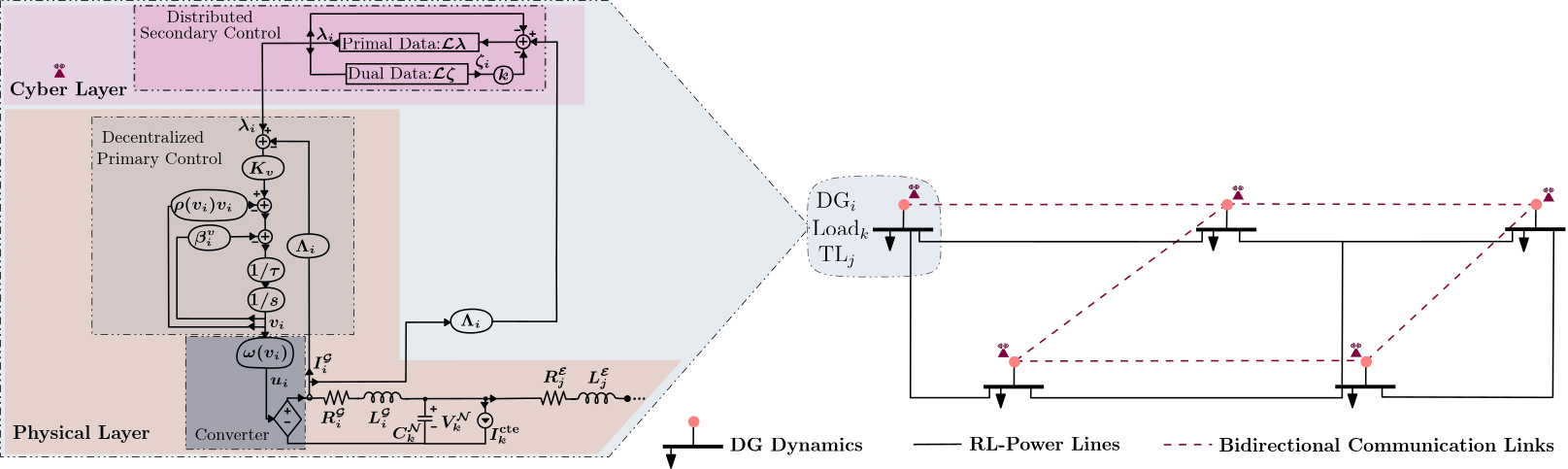}
    \captionsetup{font=small}
    \caption{Cyber Physical DC Microgrid}
    \label{Image:CPMG}
\end{figure*}

The DC microgrid architecture is inherently multilayered, consisting of a physical layer and a cyber layer as illustrated in Fig.~\ref{Image:CPMG}. In the physical layer, the electrical components (agents and loads) are interconnected through RL-modeled power lines and regulated by decentralized inner-loop controllers, while the cyber layer enables information exchange among neighboring agents as part of distributed secondary outer-loop controllers. The agents are distributed generators (DGs), located close to the power-consuming loads (ZI-loads), and are effectively interfaced with the rest of the MG through voltage-controlled converters. The converters are considered equivalent zero-order models, hence, the internal voltage controller and associated dynamics are not considered in this paper. The DGs are interconnected both electrically and via distributed communication links, forming a cyber-physical system. Graph theory is used to establish the physical and virtual interconnections, see Appendix A in \cite{Babak_AC} for the precise definitions. 
Throughout the paper, we use the following notations: $\mathbb{R}^{n\times m}$ and $\mathbb{R}^n$ denote a set of $n\times m$ real matrices and $n \times 1$ real vectors, respectively. $\mathrm{col}(\cdot\cdot\cdot) \in \mathbb{R}^n$ denotes a column vector and $\mathrm{bcol}\{\cdot\cdot\cdot\}\in \mathbb{R}^n$ denotes a column vector of vectors. $\mathrm{diag}(\cdot\cdot\cdot)\in \mathbb{R}^{n\times n}$ denotes a diagonal matrix with scalar entries, and $\mathrm{bdiag}\{\cdot\cdot\cdot\}\in \mathbb{R}^{n\times n}$ denotes a diagonal matrix of vectors. $0_n\in \mathbb{R}^n$ denotes a null vector, and $0_{n\times m} \in \mathbb{R}^{n \times n}$ denotes a null matrix. $A|_\mathrm{sym}$ denotes the symmetrical part of an arbitrary square matrix $A \in \mathbb{R}^{n \times m}$, defined as $A|_\mathrm{sym}=\frac{1}{2}(A+A^\top)$, and $\mathbb{I}=\mathrm{diag}(1)\in \mathbb{R}^{n\times n}$ is the identity matrix.%Moreover, $A_{i,i}$ refers to the diagonal entities of $A$.
Given a scalar or a vector $x$, the value at the equilibrium point is indicated as $\Bar{x}$, and $\tilde x$ denotes a shifted variable where $\tilde x \triangleq x - \bar x$. Moreover, we use the block vector notation $x^b$, for any subset $b \subseteq \mathcal{G} \cup \mathcal{E} \cup \mathcal{N}$, where $x^b$ denotes the vector collecting the states associated with the indices in $b$. Note that $\mathcal{G}=\{1, 2, \cdot \cdot \cdot, n_i\}$ is the set of distributed generators, $\mathcal{E}=\{1, 2, \cdot \cdot \cdot, n_j\}$ is the set of power lines, and $\mathcal{N}=\{1, 2, \cdot \cdot \cdot, n_k\}$ is the set of power consuming loads.

\subsection{Nonlinear Nested Distributed Control Framework}
Motivated by the previously proposed distributed controllers in \cite{Poppi_J1, Poppi_J2}, this research presents a modified version of the \emph{nonlinear} control frameworks, imposing two nested control loops: a decentralized(/primary) inner-loop controller and a consensus-based secondary outer-loop controller, aiming to simultaneously achieve of the following two control objectives in steady state:

\begin{itemize}[label=\textendash] 
    \item voltage containment within pre-specified limits,
    \item proportional current sharing.
\end{itemize}
These control objectives are mathematically formulated as follows: 
%\vspace{-.1in}
\begin{equation} \label{cont_obj}
    \begin{split}
        &V_\mathrm{min} \leq V_i^\mathcal{G}(t) \leq V_\mathrm{max}, \quad \forall \; t \in \; \mathbb{R}_{\geq 0}\\
        & \lim_{t \to \infty} \left( I_i^\mathcal{G}(t)/I_i^\mathrm{rated} -I_l^\mathcal{G}(t)/I_l^\mathrm{rated}\right)=0, \quad \forall i,l \in \mathcal{G}
    \end{split}
\end{equation}

where $V_i^\mathcal{G}$ and $I_i^\mathcal{G}$ respectively are the voltage and current output of the $i$th DG with associated rated current capacity $I_i^\mathrm{rated}$, and $V_\mathrm{min}$ and $V_\mathrm{max}$ are the allowable minimum and maximum values of the voltage output.

To facilitate the achievement of these control objectives, we embed a nonlinear hyperbolic tangent function to saturate the control input delivered to the DGs, thereby containing the voltages within pre-specified limits. Moreover, for proportional current sharing, the decentralized inner-loop integrator controller is designed to regulate the currents to a desired set-point value. This ensures that all DGs actively contribute the same proportion of power relative to their rated capacities. However, when the DG voltages reach saturation, a nonlinear leakage function is activated in the inner-loop, serving as a droop control mechanism to limit current deviations relative to the set-point value. For the \textit{i}th DG, the complete control scheme is then proposed as follows:
\begin{subequations} \label{eq:Control-Layer}
\begin{align}
    &u_i=\omega(v_i) -  \Lambda_i\mu\lambda_i,\label{omega}\\
    &\omega(v_i)=V^*+\Delta \mathrm{tanh}\left(v_i/\Delta\right), \label{omega1}\\
    &\tau_{i} \Dot{v}_i=- \rho(v_i)v_i+\mathcal{K}_i^v(\lambda_i-\Lambda_{i}I_i^{\mathcal{G}})-\mathcal{B}^v_i v_i, \label{regulator}\\
    &\tau^p_{i} \Dot{\lambda}_i = \Lambda_iI_i^{\mathcal{G}}-\lambda_i - \sum_{j\in N_i} [a_{ij} (\zeta_i-\zeta_j)-\mathrm{k} a_{ij} (\lambda_j-\lambda_i)], \label{distributed_lambda}\\
    &\tau^d_{i} \Dot{\zeta}_i = \sum_{j\in N_i} a_{ij} (\lambda_i-\lambda_j)-\mathcal{B}^\zeta_i\zeta_i, \label{distributed_zeta}\\
    &\rho(v_i)=\alpha (1+0.5[\tan\mathrm{h}(\mathrm{b}(v_i-\eta v_\mathrm{pos}))-\tan\mathrm{h}(\mathrm{b}(v_i-\eta v_\mathrm{neg}))]).  \label{leakage}
\end{align}
\end{subequations}
The distributed control framework depends on three controller states where $\lambda_i$ and $\zeta_i$ are the communicated states of the cyber network with respective time constants $\tau_i^p$ and $\tau_i^d$, and $v_i$ is the decentralized controller state with an associated time constant $\tau_i$. $u_i$ is the control input, constructed as a sum of the nonlinear function $\omega(v_i)$ and $\Lambda_i\mu\lambda_i$, where $\mu>0$ is a tuning parameter. $\omega(v_i)$ is a nonlinear hyperbolic tangent function, implemented to saturate the argument of the controller if the voltages exceed the pre-specified limits. It should be noted that incorporating $\Lambda_i \mu \lambda_i$ in the primary control dynamics diminishes \emph{complete} voltage containment. Nevertheless, by selecting $\mu$ arbitrarily small, its dynamical effect can be significantly mitigated in practice. Furthermore, we define: $V_\mathrm{max}=(1+\phi) V_\mathrm{n}$ and $V_\mathrm{min}=(1-\phi)V_\mathrm{n}$, for $\phi>0$, as the pre-specified voltage limitations from the networks nominal voltage, $V_\mathrm{n}$; $V^*=\frac{1}{2}(V_\mathrm{min}+V_\mathrm{max})$ as the central point within the safe voltage range; $\Delta=\frac{1}{2}(V_\mathrm{max}-V_\mathrm{min})$ as the maximum allowed deviation from $V^*$.
 Moreover,  $\rho(v_i)$ is a nonlinear hyperbolic tangent function serving as a leakage/droop mechanism, activated when we reach voltage saturation under the given arguments; $v_\mathrm{pos}=\Delta\tanh^{-1}[(V_\mathrm{max}-v_\mathrm{tol}-V^*)/\Delta]$ and $V_\mathrm{neg}=\Delta\tanh^{-1}[(V_\mathrm{min}+v_\mathrm{tol}-V^*)/\Delta]$, where $v_\mathrm{tol}$ is the saturation tolerance, $\alpha$ is the leakage coefficient scaling the upper bound value; $\mathrm{b}$ and $\eta$ are implemented to scale the steepness of the curve. For the decentralized integral controller, we define $\mathcal{K}_i^v>0$ as the integrator gain. $\mathcal{B}^v_i$ is a leakage constant, introducing \emph{permanent} leakage in the individual generators in the inner-loop, and $\mathcal{B}^\zeta_i$ is a uniformly defined leakage term, implemented in the outer-loop. Note that these two leakage terms are introduced for stability purposes and mathematical feasibility, respectively; however, they are initially deactivated ($\mathcal{B}^v_i=0$) and tuned negligibly small ($\mathcal{B}^\zeta_i\approx 0$). Finally, $a_{ij}$ is the \emph{adjacency matrix} element, describing the communicating DGs where $N_i$ is the set of the neighboring DGs in the cyber layer, and $k$ is a positive gain.

Note that the distributed optimizer relies solely on the per-unit currents of the associated DG and does not require information about the rated capacities of neighboring DGs—an advantageous property for privacy preservation. Consequently, the output of the physical network is defined as the ratio of each DGs generated current to its rated value: $$
\Lambda_i I_i^\mathcal{G} = I_i^\mathcal{G}/I_i^\mathrm{rated}[\mathrm{p.u.}] \quad \forall i \in \mathcal{G}, \quad \text{where } \Lambda_i =1/I_i^\mathrm{rated}.$$

\vspace{-.2in}
%and lambda is then a rated value...
\subsection{Dynamical Model of Closed-Loop DC Microgrid}
Given the proposed control framework in \eqref{eq:Control-Layer}, we express the dynamics of the closed-loop DC MG in compact form as 

\vspace{-.12in}

\begin{subequations}\label{complete_sys_compact}

    \begin{align}
    L^\mathcal{G}\Dot{I}^\mathcal{G}&=\omega(v) - \Lambda^\top\mu \lambda - \beta^\mathcal{G}V^\mathcal{N}-R^\mathcal{G}I^\mathcal{G} ,\label{complete_sys_compact_a}\\
    L^\mathcal{E}\Dot{I}^\mathcal{E}&= -\beta^\mathcal{E}V^\mathcal{N}-R^\mathcal{E}I^\mathcal{E}, \label{complete_sys_compact_b}\\
    C^\mathcal{N}\Dot{V}^\mathcal{N}&=\beta^{\mathcal{E}\top}I^\mathcal{E}+\beta^{\mathcal{G}\top}I^\mathcal{G}-G^\mathrm{cte}V^\mathcal{N}-I^\mathrm{cte},  \label{complete_sys_compact_c}\\
    \tau \Dot{v}&=-\rho(v)v+\mathcal{K}_v(\lambda-\Lambda I^\mathcal{G})-\mathcal{B}_v v, \label{complete_sys_compact_d}\\
    \tau_p \Dot{\lambda} &=\Lambda I^\mathcal{G}-\lambda-\mathcal{L}\zeta-k\mathcal{L}\lambda, \label{complete_sys_compact_e}\\
    \tau_d \Dot{\zeta} &= \mathcal{L}\lambda-\mathcal{B}_\zeta\zeta,\label{complete_sys_compact_f}
    \end{align}
\end{subequations}
%where $L^\mathcal{G}$, $I^\mathcal{G}$, and $R^\mathcal{G}$ are vector containing the inductances, currenst and resistances of the DGs. $L^\mathcal{E}$, $I^\mathcal{E}$, and $R^\mathcal{E}$ are vectors containing the inductances, currents and resistances of the thansmission lines. $C^\mathcal{N}$, $V^\mathcal{N}$, $G^\mathrm{cte}$, and $I^\mathrm{cte}$ are vector containing shunt capacitances, voltages, constant conductances, and constant currents of the power-consuming loads. Moreover, $\beta^\mathcal{G}$ and $\beta^\mathcal{E}$ are the incidence matrices of the physical network, and $\mathcal{L}$ is the \emph{Laplacian} matrix,  containing the consensus properties of the communication network. All dimensions of the dynamics in \eqref{complete_sys_compact} are given in \emph{Appendix}

where $L^\mathcal{G}=\mathrm{diag}(L_i^\mathcal{G})\in \mathbb{R}^{n_i\times n_i}$, $I^\mathcal{G}=\mathrm{col}(I_i^\mathcal{G}) \in \mathbb{R}^{n_i}$, and $R^\mathcal{G}=\mathrm{diag}(R_i^\mathcal{G}) \in \mathbb{R}^{n_i\times n_i}$ respectively contain the inductances ($L_i^\mathcal{G}$), currents ($I_i^\mathcal{G}$) and resistances ($R_i^\mathcal{G}$) of the $i$th DG $\forall i \in \mathcal{G}$. $L^\mathcal{E}=\mathrm{diag}(L_j^\mathcal{E})\in \mathbb{R}^{n_j\times n_j}$, $I^\mathcal{E}=\mathrm{col}(I_j^\mathcal{E}) \in \mathbb{R}^{n_j}$, and $R^\mathcal{E}=\mathrm{diag}(R_j^\mathcal{E}) \in \mathbb{R}^{n_j\times n_j}$ respectively contain the inductances ($L_j^\mathcal{E}$), currents ($I_j^\mathcal{E}$), and resistances ($R_j^\mathcal{E}$) of the $j$th power line $\forall j \in \mathcal{E}$. $C^\mathcal{N}=\mathrm{diag}(C_k^\mathcal{N})\in \mathbb{R}^{n_k\times n_k}$, $V^\mathcal{N}=\mathrm{col}(V_k^\mathcal{N}) \in \mathbb{R}^{n_k}$, $G^\mathrm{cte}=\mathrm{diag}(G_k^\mathrm{cte}) \in \mathbb{R}^{n_k\times n_k}$, and $I^\mathrm{cte}=\mathrm{col}(I_k^\mathrm{cte}) \in \mathbb{R}^{n_k}$ respectively contain the shunt capacitances ($C_k^\mathcal{N}$), voltages ($V_k^\mathcal{N}$), constant conductances ($G_k^\mathrm{cte}$), and constant currents ($I_k^\mathrm{cte} $) of the $k$th power-consuming loads $\forall k \in \mathcal{N}$. $\beta^\mathcal{G}=[b_{ik}^\mathcal{G}]\in \mathbb{R}^{n_i\times n_k}$ and $\beta^\mathcal{E}=[b_{jk}^\mathcal{E}]\in \mathbb{R}^{n_j\times n_k}$ are the incidence matrices of the physical network, where the elements $b_{ik}^\mathcal{G}$ and $b_{jk}^\mathcal{E}$ are either $-1$, $1$ or $0$--characterizing arbitrary current flows within the network. Furthermore, we define $\omega(v)=\mathrm{col}(\omega(v_i)) \in \mathbb{R}^{n_i}$, $v =\mathrm{col}(v^c_{i}) \in \mathbb{R}^{n_i}$, $\tau=\mathrm{diag}(\tau_i) \in \mathbb{R}^{n_i\times n_i}$, $\rho(v)=\mathrm{diag}(\rho(v_i))\in \mathbb{R}^{n_i \times n_i}$, $\sigma(\lambda)=\mathrm{col}(\sigma_i(\lambda_i))\in \mathbb{R}^{n_i}$, $\mathcal{K}_v=\mathrm{diag}(\mathcal{K}_i^v)\in \mathbb{R}^{n_i\times n_i}$, $\mathcal{B}_v=\mathrm{diag}(\mathcal{B}^v_i) \in \mathbb{R}^{n_i\times n_i}$, $\lambda=\mathrm{col}(\lambda_i) \in \mathbb{R}^{n_i}$, $\Lambda=\mathrm{diag}(1/I^\mathrm{rated}_i) \in \mathbb{R}^{n_i\times n_i}$, $\zeta=\mathrm{col}(\zeta_i) \in \mathbb{R}^{n_i}$, $\mathcal{B}_\zeta=\mathrm{diag}(\mathcal{B}^\zeta_i) \in \mathbb{R}^{n_i\times n_i}$, $\tau_p=\mathrm{diag}(\tau^p_i) \in \mathbb{R}^{n_i \times n_i}$, $\tau_d=\mathrm{diag}(\tau^d_i) \in \mathbb{R}^{n_i\times n_i}$. Finally, $\mathcal{L}=[l_{ij}]\in \mathbb{R}^{n_i \times n_i}$ is the \emph{Laplacian} matrix,  containing the consensus properties of the communication network.

\begin{assumption}
\label{ass_Laplacian}
    It is assumed that the communication network (represented by the Laplacian matrix $\mathcal{L}$) is strongly connected and undirected. 
\end{assumption}
\vspace{-.25in}
 
\section{System Steady State and Stability} \label{stability}

We aim to prove that the equilibrium of \eqref{complete_sys_compact} is globally exponentially stable using singular perturbation theory \cite{Khalil}. We apply singular perturbation theory to obtain \emph{scalable} stability conditions to the closed-loop DC MG system in \eqref{complete_sys_compact}, and apply Lyapunov theory to obtain a G.E.S. certificate. To facilitate applying SPT we assume sufficient time-scale separation, where \eqref{complete_sys_compact_a}-\eqref{complete_sys_compact_c},\eqref{complete_sys_compact_e}-\eqref{complete_sys_compact_f} are considered as the fast dynamics, and \eqref{complete_sys_compact_d} is considered the slow dynamics under following assumption.
\begin{assumption}(Time-Scale Separation)
\label{Asump_time_Const}
    The decentralized inner-loop integrator is characterized by the largest time constant of the closed-loop microgrid, such that $\tau\gg\varepsilon$, where $\varepsilon$ denotes the largest time constant of the fast dynamics.
    %$\tau > \{\tau_p, \tau_d, \frac{L^\mathcal{G}}{R^\mathcal{G}}, \frac{L^\mathcal{E}}{R^\mathcal{E}}, \frac{C^\mathcal{N}}{G^\mathrm{cte}}\}$ with $\varepsilon=\tau$ where $\tau$ is the slowest time constant of the fast dynamics.
\end{assumption}
\begin{Remark}
    Given the commonly adopted hierarchical control architectures in DC microgrids, it is reasonable to assume that the outer-loop dynamics (distributed controller) operates at a slower time-scale than the inner-loop dynamics (decentralized integrator and electrical dynamics). However, we propose a less traditional time-scale separation, as the decentralized integrator is assumed to operate at a slower time-scale than both the outer-loop dynamics and the electrical system dynamics.
\end{Remark}

\noindent Let \emph{Assumption} \ref{Asump_time_Const} hold, and let the two time-scaled separated systems be defined as
\vspace{-.1in}
\begin{align} \label{two_separated}
    \Dot{v}=s(v, z) \quad \text{and} \quad 
        \varepsilon \Dot{z}=f(v, z)
\end{align}
where $s(\cdot)$ and $f(\cdot)$ respectively denotes the slow and fast system dynamics, given $v=\mathrm{col}(v_i)\in \mathbb{R}^{n_\mathcal{S}}$ and $z=\mathrm{bcol}\{I^\mathcal{G}, I^\mathcal{E}, V^\mathcal{N}, \lambda, \zeta\} \in \mathbb{R}^{n_\mathcal{F}}$, with $n_\mathcal{F}=3\times  n_i+n_j+n_k$ and $n_\mathcal{S}=n_i$. 
%\vspace{-.15in}

\subsection{Equilibrium Analysis}\label{sec:SS}
We derive the steady-state equations of \eqref{complete_sys_compact} by expressing $\Dot{v}=0_{n_\mathcal{S}}$, and $\dot z = 0_{n_\mathcal{F}}$. Hence, any steady state needs to satisfy
\begin{subequations}
\begin{IEEEeqnarray}{lCl}
        0_{n_\mathcal{F}}=f(\Bar{I}^\mathcal{G}, \Bar{I}^\mathcal{E}, \Bar{V}^\mathcal{N}, \bar v, \bar \lambda, \bar \zeta),\label{ST_P} \nonumber\\
        0_{n_\mathcal{S}}=s(\Bar{v}, \Bar{\lambda},\Bar{I}^\mathcal{G}), \label{ST_C} \nonumber
\end{IEEEeqnarray}
\end{subequations}
Let \emph{Assumption} \ref{ass_Laplacian} hold, and let the influence of $\mathcal{B}_\zeta$ be negligible. By leveraging the property of the Laplacian matrix where $\mathcal{L}1_{n_i}=0_{n_i}$, the steady state operations of \eqref{complete_sys_compact_f} guarantee consensus among the communicating agents, such that $\Bar{\lambda}=\lambda_s1_{n_i}$ for $\lambda_s$ being the singular consensus value and $1_{n_i}=\mathrm{col}(1)\in \mathbb{R}^{n_i}$. Subsequently, considering the steady state of \eqref{complete_sys_compact_d} in unsaturated operations; i.e., inactive $\rho(v)$, and with $\mathcal{B}_v\approx 0$, $\Lambda\Bar{I}^\mathcal{G}$ is forced to be equal to $\bar \lambda$. Accordingly, steady state operations of \eqref{complete_sys_compact_e} forces $\Bar{\zeta}=\zeta_s1_{n_i}$ due to the properties of the Laplacian matrix, ensuring consensus among the DGs in regards to the second communicated state variable $\zeta$. The steady state of the electrical network \eqref{complete_sys_compact_a}-\eqref{complete_sys_compact_c} is consequently uniquely defined. 

\subsubsection{Optimal Steady State Conditions} \label{sec_SS_opt}
To guarantee \emph{optimal} operation, it is necessary to verify that the attained steady-state equilibrium point compiles with the optimality conditions-- in which the two control objectives in \eqref{cont_obj} are satisfied. Voltage containment is dynamically guaranteed at all times through the hyperbolic tangent function $\omega(v)$. For proportional current sharing, the decentralized inner-loop controller is designed to regulate the DG currents (in [p.u.]) to an \emph{optimal} and uniformly defined set-point value. Following the control design strategy presented in \cite{Babak_Set_Point}, this \emph{optimal} value is established by the solution that minimizes the cost function of the \emph{Distributed Multi-Variable} optimization problem -- introduced in Section $\mathbb{II}$.B of \cite{Babak_Set_Point}. Applied to our system, this optimization algorithm is formulated as follows:
%\vspace{-.05in}
\begin{align}\label{opt_prob}
\begin{split}
&\min_{(\bar{\lambda}_1, \ldots, \bar{\lambda}_{n_i})} \quad  \frac{1}{2} \sum_{i=1}^{n_i} (\bar{\lambda}_i - \Lambda\bar{I}^\mathcal{G}_i)^2, \\
&\text{subject to} \quad  \bar{x}_i = \sum_{j \in N_i} a_{ij}(\bar{\lambda}_i - \bar{\lambda}_j) = 0, \quad \forall i, 
\end{split}
\end{align}
\vspace{-.25in}

where the constraint implies that the communicating agents attains consensus when $\bar \lambda_i=\bar \lambda_j \forall i,j \in N_i$. The cost function to be minimized is quadratic, and the constraint is linear, thus, Slater’s condition is satisfied. Accordingly, we formulate the following Lagrangian in \eqref{opt_lag}, from which the primal and dual Karush–Kuhn–Tucker (KKT) conditions are derived, defining our steady state optimality conditions.
\vspace{-.07in}
\begin{align}\label{opt_lag}
    \mathbb{L}(\bar{\lambda}_i, \bar{\zeta}_i)
= \frac{1}{2} \sum_{i=1}^{n_i} (\bar{\lambda}_i - \Lambda\bar{I}^\mathcal{G}_i)^2
+ \sum_{i=1}^{n_i} \bar{\zeta}_i \bar{x}_i, 
\quad \forall i \in \mathcal{G}.
\end{align}
\vspace{-.07in}

$\bar{\zeta}_i$ denotes the Lagrange multiplier associated with $\bar{x}_i$. The variables $\bar{\lambda}$ and $\bar{\zeta}$ are then considered optimal if they satisfy the following KKT conditions:
\vspace{-.1in}
\begin{align*}
0 &= \bar{\lambda}_i - \Lambda
\bar{I}^\mathcal{G}_i + \sum_{j \in N_i} a_{ij} (\bar{\zeta}_i - \bar{\zeta}_j), \\
0 &= \bar{x}_i = \sum_{j \in N_i} a_{ij} (\bar{\lambda}_i - \bar{\lambda}_j). 
\end{align*}
\vspace{-.1in}

When the KKT conditions are expressed in the compact form as in \eqref{KKT_comp}, it becomes evident that the steady-state operation analyzed in the previous section satisfies them when saturation is avoided--i.e., when the \emph{non-permanent} leakage $\rho(v)$ remains inactive--and without any \emph{permanent} leakage term. In this case, the proposed controller in \eqref{eq:Control-Layer} simultaneously ensures \emph{optimal} proportional current sharing and voltage containment, provided that all neighboring DGs collaboratively determine the \emph{optimal} set-point value $\bar{\lambda}$ such that $\Lambda\bar{I}^\mathcal{G} = \bar{\lambda}$ at equilibrium. However, in the case of leakage activation, the MG is steered to another optimum that fails to meet the \emph{optimal} KKT conditions.
\vspace{-.05in}
\begin{align}\label{KKT_comp}
    \begin{split}
    0_{n_i}&=\Lambda \bar I^\mathcal{G}-\bar \lambda - \mathcal{L}\bar \zeta \\
    0_{n_i} &= \mathcal{L}\bar \lambda 
    \end{split}
\end{align}
\vspace{-.25in}

\subsection{Singular Perturbation Theory} \label{sec:GES}

To accommodate the nested control loops, we first differentiate the dynamical system into \emph{slow/fast} models (given in \emph{Assumption} \ref{Asump_time_Const}). Following the methodology in \cite{Khalil}, we express the system dynamics in \eqref{complete_sys_compact} as a \emph{singular perturbation problem} under a stretched time-scale ($\mathbf{t}$), partitioning the problem into the \emph{reduced}- and \emph{boundary layer} subsystems. By proposing two separate Lyapunov candidates for the two time-scaled systems, we aim to prove G.E.S. of the individual systems under some stability bounds. This facilitates finding a composite Lyapunov function for the singularly perturbed system as a weighted sum of the two Lyapunov candidates, used to conclude on G.E.S. of the \emph{full} system. 

%\footnote{For this time-scale separation, the dynamics in $\dot v$ exclusively constitute the slow system, and $v$ is considered the only slow variable.}
%

\begin{Prop5} (Singular Perturbed Problem) \label{teorem_SPT}
Consider the closed loop dynamics in \eqref{complete_sys_compact}, and let
\begin{align*}
    %\mathbf{z}&\triangleq y+\mathrm{h}(v)\\
    \Omega(\tilde v) &\triangleq \omega(v)-\omega(\bar v) \\
    \Gamma(\tilde v) &\triangleq \rho(v)v - \rho (\bar v)\bar v,\\
    \mathcal{Q}_\mathcal{F} &\triangleq\mathrm{bdiag}\{L^\mathcal{G},L^\mathcal{E},C^\mathcal{N}, \tau_p, \tau_d\}>0\\
    \mathcal{P}_\mathcal{F} &\triangleq \mathrm{bdiag}\{R^\mathcal{G},R^\mathcal{E},G^\mathrm{cte}, (\mathbb{I}+k\mathcal{L}), \mathcal{B}_\zeta \}>0, \\
    \mathbf{J}_\mathcal{F}&\triangleq\begin{bmatrix}
        \mathbf{0}_{n_i\times n_i} & \mathbf{0}_{n_i\times n_j} & -\beta^\mathcal{G} & -\Lambda^\top\mu & \mathbf{0}_{n_i\times n_i}\\
        \mathbf{0}_{n_j\times n_i} & \mathbf{0}_{n_j\times n_j} & -\beta^\mathcal{E} & \mathbf{0}_{n_j\times n_i} & \mathbf{0}_{n_j\times n_i}\\
        \beta^{\mathcal{G}\top} & \beta^{\mathcal{E}\top} & \mathbf{0}_{n_k\times n_k}& \mathbf{0}_{n_k\times n_i} & \mathbf{0}_{n_k\times n_i}\\
        \Lambda & \mathbf{0}_{n_i\times n_j} & \mathbf{0}_{n_i\times n_k} & \mathbf{0}_{n_i\times n_i} & -\mathcal{L}\\
        \mathbf{0}_{n_i\times n_i} & \mathbf{0}_{n_i\times n_j} & \mathbf{0}_{n_i\times n_k} & \mathcal{L} & \mathbf{0}_{n_i\times n_i}        
    \end{bmatrix}
\end{align*}
where $\mathrm{h}(v)=\mathrm{bcol}\{\mathrm{h}^{I^\mathcal{G}}_i(v), \mathrm{h}_j^{I^\mathcal{E}}(v), \mathrm{h}^{V^\mathcal{N}}_k (v), \mathrm{h}^{\lambda}_i(v), \mathrm{h}^{\zeta}_i(v)\}, \forall i \in \mathcal{G}, \forall j \in \mathcal{E}, \forall k \in \mathcal{N}$ is the unique solution of $\varepsilon \dot z = f(v, z)$ for $\varepsilon \approx 0$; $\tilde y\triangleq \mathrm{bcol}\{\tilde z - \mathrm{H}(\tilde v)\}$; $\mathrm{H}(\tilde v)\triangleq \mathrm{bcol}\{ \hat{\mathrm{H}}(\tilde v)-\mathrm{h}(\bar v)\}$; $\hat{\mathrm{H}}(\tilde v)\triangleq \mathrm{bcol}\{\mathrm{h}(\tilde v+\bar v)\}$. Suppose that $\Omega(\tilde v)$, $\Gamma(\tilde v)$ are Lipschitz and element-wise strictly monotonically increasing sigmoid functions in $\tilde v$. 
The two separated time-scaled systems in \eqref{two_separated} can then be expressed as the singular perturbed problem under the stretched timescale $\mathbf{t}=(t/\varepsilon)$, where $\hat{s}(\tilde v, \mathrm{H}(\tilde v))$ and $\hat{f}(\tilde v, \tilde y+\mathrm{H}(\tilde v))$ represent the reduced and boundary layer systems, respectively. 

\vspace{-.2in}
    \begin{subequations} \label{SP_Compact_Inc}
        \begin{align}
            &\hat{s}(\tilde v, \mathrm{H}(\tilde v)) : \begin{cases}
             \tau  \Dot{ \tilde v} = -\Gamma(\tilde v) + \mathcal{K}_v[\mathrm{H}^\lambda(\tilde v) - \Lambda \mathrm{H}^{I^\mathcal{G}}(\tilde v) ]-\mathcal{B}_vv,
             \end{cases} \label{reduced}\\
            &\hat{f}(\tilde v, \tilde y+\mathrm{H}(\tilde v)):\begin{cases}
                \mathcal{Q}_\mathcal{F} \partial \tilde y/\partial \mathbf{t} = [\mathbf{J}_\mathcal{F}-\mathcal{P}_\mathcal{F}] (\tilde y + \mathrm{H}(\tilde v)) \label{boundary}
            \end{cases} 
        \end{align}
    \end{subequations}
    \vspace{-.2in}

\end{Prop5}

\begin{proof}

First, we consider $\varepsilon$ to be defined by the slowest time constant of the fast system, such that the velocity of $\Dot{z}\propto (1/\varepsilon)$ behaves instantaneously fast for a sufficiently small $\varepsilon$. Secondly, for $\varepsilon \approx 0$, the fast system in \eqref{complete_sys_compact_a}-\eqref{complete_sys_compact_c}, \eqref{complete_sys_compact_e}-\eqref{complete_sys_compact_f} quickly reaches a \emph{quasi-steady state} given the instantaneous fast dynamics, with $\mathrm{h}(v) \in \mathbb{R}^{n_\mathcal{F}}$ defined in \emph{Theorem} \ref{teorem_SPT}.
 
    \vspace{-.15in}
    \begin{IEEEeqnarray}{lCl}\label{h(x)}
   0_{n_\mathcal{F}}=\begin{bmatrix}
        \omega(v)-\Lambda^\top \mathrm{h}^\lambda(v)-\beta^\mathcal{G}\mathrm{h}^{V^\mathcal{N}}(v)-R^\mathcal{G}\mathrm{h}^{I^\mathcal{G}}(v)\\
        -\beta^\mathcal{E}\mathrm{h}^{V^\mathcal{N}}(v)-R^\mathcal{E}\mathrm{h}^{I^\mathcal{E}}(v)\\
        \beta^{\mathcal{G}\top}\mathrm{h}^{I^\mathcal{G}}(v)+\beta^{\mathcal{E}\top} \mathrm{h}^{I^\mathcal{E}}(v)-G^\mathrm{cte}\mathrm{h}^{V^\mathcal{N}}(v)-I^\mathrm{cte}\\
        \Lambda \mathrm{h}^{I^\mathcal{G}}(v)-\mathrm{h}^\lambda(v)-\mathcal{L}\mathrm{h}^\zeta(v)-k\mathcal{L}\mathrm{h}^\lambda(v)\\
        \mathcal{L}\mathrm{h}^\lambda(v)-\mathcal{B}_\zeta \mathrm{h}^\zeta (v)
    \end{bmatrix}
    \end{IEEEeqnarray}
    %\vspace{-.15in}
    
    We further define $y=\mathrm{bcol}\{y_i^{I^\mathcal{G}}, y_j^{I^\mathcal{E}}, y_k^{v^\mathcal{N}}, y_i^\lambda, y_i^\zeta\}$ as the error between the actual fast dynamics and the quasi-steady state: $y^b\triangleq \mathrm{bcol} \{z^b-\mathrm{h}^b(v)\}, \;  \forall b\in \{I^\mathcal{G}, I^\mathcal{E}, V^\mathcal{N}, \lambda, \zeta\}$. Moreover, to facilitate applying the Lyapunov theory on the two time-separated systems, we express the system states using their incremental variables $\tilde z= z - \bar z$, $\Tilde{y}=y-\Bar{y}$ and $\Tilde{v}=v-\Bar{v}$. Thus, we express the incremental \emph{real} fast dynamics as $\Tilde{\mathbf{z}} \triangleq \tilde y + \mathrm{h}(\tilde v + \bar v) - \mathrm{h}(\bar v)=\tilde y + \mathrm{H}(\tilde v)$ where we use the fact that $\bar z =\mathrm{h}(\bar v)$, such that $\bar y=0$. Following \emph{Assumption} \ref{Asump_time_Const}, we express the fast and slow dynamical models as
    \vspace{-.05in}
    \begin{subequations} \label{slow/fast_Compact_Inc}
        \begin{align}
            &s(\tilde v, \tilde z) : \begin{cases}
             \tau \Dot{ \tilde v} = -\Gamma(\tilde v) + \mathcal{K}_v[\Tilde{\mathbf{z}}_\lambda - \Lambda \Tilde{\mathbf{z}}_{I^\mathcal{G}} ]-\mathcal{B}_v\tilde v,
             \end{cases} \label{Slow_compact_inc}\\
            &f(\tilde v, \tilde z) :  \begin{cases}
                \mathcal{Q}_\mathcal{F} \Dot{\Tilde{\mathbf{z}}}_\mathcal{F} = [\mathbf{J}_\mathcal{F}-\mathcal{P}_\mathcal{F}] \Tilde{\mathbf{z}}_\mathcal{F} + \kappa_1 \Omega(\tilde v),\label{fast_compact_inc}
            \end{cases} 
        \end{align}
    \end{subequations}  
    
    where 
    %$\kappa_1 \triangleq \mathrm{bdiag}\{\mathbf{1}_{n_1 \times n_i}, \mathbf{0}_{nj+nk+2ni \times nj+nk+2ni  }\}$\\
    $\kappa_1 \triangleq \mathrm{bdiag}\{\mathbf{1}_{n_i \times n_i},\mathbf{0}_{n_j\times n_j}, \mathbf{0}_{n_k\times n_k}, \mathbf{0}_{n_i\times n_i}, \mathbf{0}_{n_i\times n_i}\}$.

     For compactness, we have defined the following for the fast model; the physical inertia matrix $\mathcal{Q}_\mathcal{F}$; the physical resistance matrix $\mathcal{P}_\mathcal{F}$; and the interconnection matrix $\mathbf{J}_\mathcal{F}$ whose structures are defined in \emph{Theorem} \ref{teorem_SPT}. Moreover, for mathematical purposes, we have introduced an additional leakage constant $\mathcal{B}_\zeta$, ensuring $\mathcal{P}_\mathcal{F}$ to be a fully diagonal and positive-definite matrix. Note that $\mathcal{B}_\zeta$ does not need to compensate for any off-diagonal elements in the electrical resistance matrix. Hence, we choose the value to be arbitrarily small not to hinder the capability of achieving consensus for an optimal value of $\lambda$. 
    The system in \eqref{slow/fast_Compact_Inc} is then expressed as the \emph{singular perturbation problem} in \eqref{SP_Compact_Inc}, partitioned into the reduced system \eqref{reduced} and the boundary layer system \eqref{boundary} divided under the stretched timescale $\mathbf{t}=(t/\varepsilon)$ where $t$ is the time when $\varepsilon \approx 0$. Thus, the slow model \eqref{Slow_compact_inc} will instantaneously achieve a quasi-steady state as the fast dynamics are considered instantaneously fast, such that $\tilde y\approx 0$ in the reduced system \eqref{reduced}. Note that even though the boundary layer system depends on both slow ($\tilde v$) and fast ($\tilde y$) dynamics ($\hat f(\tilde v, \tilde y+\mathrm{H}(\tilde v)$), the slow states entering the boundary layer system are considered frozen variables due to sufficient time-scale separation. Thus, the dynamical response in the fast system--expressed with incremental states-- only depends on the change in the error between the real fast dynamics and the instantaneous fast dynamics ($\tilde y$). Hence, all dynamics concerning a change in solely the slow states are disregarded in the boundary layer system in \eqref{boundary}.
\end{proof}

\subsection{Lyapunov Stability Analysis} \label{sec_lyap} 
\begin{Prop5} (Global Exponential Stability) \label{theorem_GES} 
Consider the singular perturbed problem in \eqref{SP_Compact_Inc}, and let $\mu>0$ and
\begin{align*}
    \mathcal{Q}^* &\triangleq\mathrm{bdiag}\{\frac{L^\mathcal{G}}{\mu},\frac{L^\mathcal{E}}{\mu},\frac{C^\mathcal{N}}{\mu}, \tau_p, \tau_d\}>0,\\
    \mathcal{J}^*&\triangleq\mathcal{Q}^*\mathcal{Q}_\mathcal{F}^{-1}\mathbf{J}_\mathcal{F}\\
    &=\begin{bmatrix}
        \mathbf{0}_{n_i\times n_i} & \mathbf{0}_{n_i\times n_j} & -\frac{\beta^\mathcal{G}}{\mu} & -\Lambda^\top & \mathbf{0}_{n_i\times n_i}\\
        \mathbf{0}_{n_j\times n_i} & \mathbf{0}_{n_j\times n_j} & -\frac{\beta^\mathcal{E}}{\mu} & \mathbf{0}_{n_j\times n_i} & \mathbf{0}_{n_j\times n_i}\\
        \frac{\beta^{\mathcal{G}\top}}{\mu} & \frac{\beta^{\mathcal{E}\top}}{\mu} & \mathbf{0}_{n_k\times n_k}& \mathbf{0}_{n_k\times n_i} & \mathbf{0}_{n_k\times n_i}\\
        \Lambda & \mathbf{0}_{n_i\times n_j} & \mathbf{0}_{n_i\times n_k} & \mathbf{0}_{n_i\times n_i} & -\mathcal{L}\\
        \mathbf{0}_{n_i\times n_i} & \mathbf{0}_{n_i\times n_j} & \mathbf{0}_{n_i\times n_k} & \mathcal{L} & \mathbf{0}_{n_i\times n_i}        
    \end{bmatrix},\\
    \mathcal{P}^*&\triangleq\mathcal{Q}^*\mathcal{Q}_\mathcal{F}^{-1}\mathcal{P}_\mathcal{F}=\mathrm{bdiag}\left\{\frac{R^\mathcal{G}}{\mu}, \frac{R^\mathcal{E}}{\mu}, \frac{C^\mathrm{cte}}{\mu}, (\mathbb{I}+k\mathcal{L}), \mathcal{B}_\zeta \right\}  >0,\\
    \mathcal{Z}( v)&\triangleq\Gamma(v)-\mathcal{K}_v[\mathrm{H}^\lambda(v)-\Lambda \mathrm{H}^{I^\mathcal{G}}(v)]+\mathcal{B}_vv.
\end{align*}
If $\mathcal{Z}(v)$ is strictly monotonically increasing; that is, \vspace{-.08in}
\begin{align} \label{monotonicity_of_M}
    \frac{1}{2}[\nabla^\top \mathcal{Z}(v) + \nabla \mathcal{Z}(v)^\top] >0,
\end{align}
then, there exists $\varepsilon^*>0$ such that for all $\tau \gg \varepsilon^*$ the system in \eqref{complete_sys_compact} is globally exponentially stable under the following bound 
\begin{align} \label{monton}
    (v-\bar v)^\top [\mathcal{Z}(v)-\mathcal{Z}(\bar v)] \leq \gamma ||v||^2, \quad \gamma>0.
\end{align}

\end{Prop5}
%For the boundary layer system, we use the following definitions; the reduced error state vector $\tilde y_f\triangleq\mathrm{bcol}\{\tilde y_1, \tilde y_2, \tilde y_3\}$ where $\mathrm{H}_f(\tilde x) \triangleq \mathrm{bcol}\{\mathrm{H}1(\tilde x), \mathrm{H}2(\tilde x), \mathrm{H}3(\tilde x)\}$. 

    \begin{proof}
    Considering the singularly perturbed system \eqref{SP_Compact_Inc}, we define the subsequent Lyapunov candidates for the boundary layer system and the reduced system, to ensure global exponential stability of the two time-separated systems. First, we express the Lyapunov candidate as follows:
    \begin{align}\label{Lyap_fast}
    V_\mathcal{F}(\tilde v, \tilde y)&=\frac{1}{2}[\tilde y+\mathrm{H}(\tilde v)]^\top \mathcal{Q}^*[\tilde y+\mathrm{H}(\tilde v)] >0
    \end{align}
    Taking the time derivative of the Lyapunov function gives
    \begin{align}\label{derive_Lyap_Fast}
    \Dot{V}_\mathcal{F}(\tilde v, \tilde y)&=\nabla^\top V_\mathcal{F}(\tilde v,  \tilde y)\Dot{\Tilde{y}}  \nonumber \\
        &= [\tilde y+\mathrm{H}(\tilde v)]^\top \mathcal{Q}^*\mathcal{Q}_\mathcal{F}^{-1}[\mathbf{J}_\mathcal{F}-\mathcal{P}_\mathcal{F}][\tilde y+\mathrm{H}(\tilde v)] \nonumber \\
        &= [\tilde y+\mathrm{H}(\tilde v)]^\top [\mathcal{J}^*-\mathcal{P}^*][\tilde y+\mathrm{H}(\tilde v)] \nonumber \\
        &= - [\tilde y+\mathrm{H}(\tilde v)]^\top\mathcal{P}^* [\tilde y+\mathrm{H}(\tilde v)]\leq 0,
    \end{align}
where we have used the skew-symmetric properties of $\mathcal{J}^*$ in the third equality. The last inequality of \eqref{derive_Lyap_Fast} is obtained by noticing that the first term is negative definite as $\mathcal{P}^* >0$ for $\mu>0$.
For the reduced system, we define $\mathcal{Q}_\mathcal{S}=\tau>0$, and the following Lyapunov candidate and its time derivative 
\begin{align}
    W_\mathcal{S}(\tilde v)&=\frac{1}{2}\tilde v^\top \mathcal{Q}_\mathcal{S} \tilde v>0\\
    \dot W_\mathcal{S} (\tilde v)&= \nabla^\top W_\mathcal{S}(\tilde v) \Dot{\tilde v}\nonumber \\
    &= -\tilde v^\top \Gamma(\tilde v) + \tilde v^\top \mathcal{K}_v[\mathrm{H}^\lambda(\tilde v)-\Lambda \mathrm{H}^{I^\mathcal{G}}(\tilde v)]-\tilde v^\top \mathcal{B}_v \tilde v \nonumber \\
    &= - (v-\bar v)^\top [\mathcal{Z}(v)-\mathcal{Z}(\bar v)] \leq 0.
\end{align}
By assumption $\mathcal{Z}(v)$ is strictly monotonically increasing and conclude that the reduced system is G.E.S. under the following bound
\begin{align}
    (v-\bar v)^\top [\mathcal{Z}(v)-\mathcal{Z}(\bar v)] \leq \gamma ||v||^2
\end{align} 
for some scalar $\gamma > 0$.

%%%%%%%%%%%%
   
    Since both $V_\mathcal{F}(\tilde v, \tilde y)$ and $W_\mathcal{S}(\tilde{v})$ are positive definite and radially unbounded, we conclude G.E.S. of the reduced and the boundary layer system. Hence, the singularly perturbed system, given in \eqref{SP_Compact_Inc}, satisfies all conditions given in \emph{Theorem} 11.4 in \cite{Khalil} and \emph{Theorem} \ref{theorem_GES} and the system in \eqref{complete_sys_compact} is globally exponentially stable as there exists a $\varepsilon^*>0$ for all $\tau \gg \varepsilon^*$. 
\end{proof}
The stability bound in \eqref{monton} highly depends on $\mathcal{Z}(v)$ to be strictly monotonically increasing. To verify this assumption, we apply the \emph{Geršgorin Circle Theorem} to guarantee (positive) diagonal dominance of $\frac{\partial\mathcal{Z}(v)}{\partial v}|_{\mathrm{sym}}$ under careful tuning. 
\begin{assumption}\label{ass:Monoton}
    Following the definition in \eqref{monotonicity_of_M}, we assume $\mathcal{Z}(v)$ to be a strictly monotonically increasing function, satisfying
    \begin{align}\label{sym_z(v)}
        \frac{\partial\mathcal{Z}(v)}{\partial v}|_{\mathrm{sym}}> 0.
    \end{align}
\end{assumption}

\begin{Prop4} \label{cor:Gersh}

    Let Assumption \ref{ass:Monoton} hold. Then, the Theorem 6.1.1. (Geršgorin) in \cite{Gersh} guarantees that the resulting Geršgorin discs lie entirely in the right half-plane for all $v_i, \forall i \in \mathbb{R}^\mathcal{G}$.

\end{Prop4}
\begin{proof}
In \eqref{z_matrix}, we compute the Jacobian of $\mathcal{Z}(v)$ and assess its symmetrical entities. Note that $\rho(v)$ is previously defined as a strictly monotonically increasing function in \emph{Theorem} \ref{teorem_SPT}, appearing only on the diagonal of the matrix. Similarly, by construction, the \emph{permanent} leakage $\mathcal{B}_v$ also appears solely on the diagonal. Therefore, we let $\mathrm{R}(v_i)\triangleq \Gamma(v_i)+\mathcal{B}^v_{i,i}v_i$. 
    \begin{align} \label{z_matrix}
        &\frac{\partial \mathcal{Z}(v)}{\partial v}|_\mathrm{sym}=\\
         &\begin{bmatrix}
        \frac{\partial(\mathrm{R}(v_1)-\mathcal{K}_v[\mathrm{H}^{\lambda}_1(v)-\Lambda \mathrm{H}^{I^\mathcal{G}}_1(v)])}{\partial v_1} & \cdots & \frac{-\partial(\mathcal{K}_v[\mathrm{H}^{\lambda}_1(v)-\Lambda \mathrm{H}^{I^\mathcal{G}}_1(v)])}{\partial v_n}  \\
        \vdots & \ddots & \vdots\\
        \frac{-\partial(\mathcal{K}_v[\mathrm{H}^{\lambda}_n(v)-\Lambda \mathrm{H}^{I^\mathcal{G}}_n(v)])}{\partial v_1} & \cdots & \frac{\partial(\mathrm{R}(v_n)-\mathcal{K}_v[\mathrm{H}^{\lambda}_n(v)-\Lambda \mathrm{H}^{I^\mathcal{G}}_n(v)])}{\partial v_n}
        \end{bmatrix} \nonumber
    \end{align}
    %We apply the Geršgorin Circle theorem on the matrix in \eqref{z_matrix} to guarantee that all eigenvalues are in the right half plane, ensuring the matrix to be positive definite. 

    %%%%
    Considering the matrix above, we adopt Theorem 6.1.1. (Geršgorin) in \cite{Gersh} to obtain the Geršgorin \emph{discs}, which provide estimated bounds on the locations of the eigenvalues. According to the theorem, the eigenvalues of each row lies within a disc, with its center at its diagonal element, and with a radius equal to the sum of the absolute values of the off-diagonal elements in the corresponding row. To guarantee that the matrix is positive definite, it suffices to ensure that, for each row, the center is greater than the radius, so that the corresponding discs lies entirely in the right half-plane. When this condition holds for every row, the union of all Geršgorin discs remains in the right half-plane, thereby ensuring that all eigenvalues are positive and the matrix is positive definite. 
\end{proof}

\begin{figure*}[!t]
    \centering
    \includegraphics[width=\textwidth]{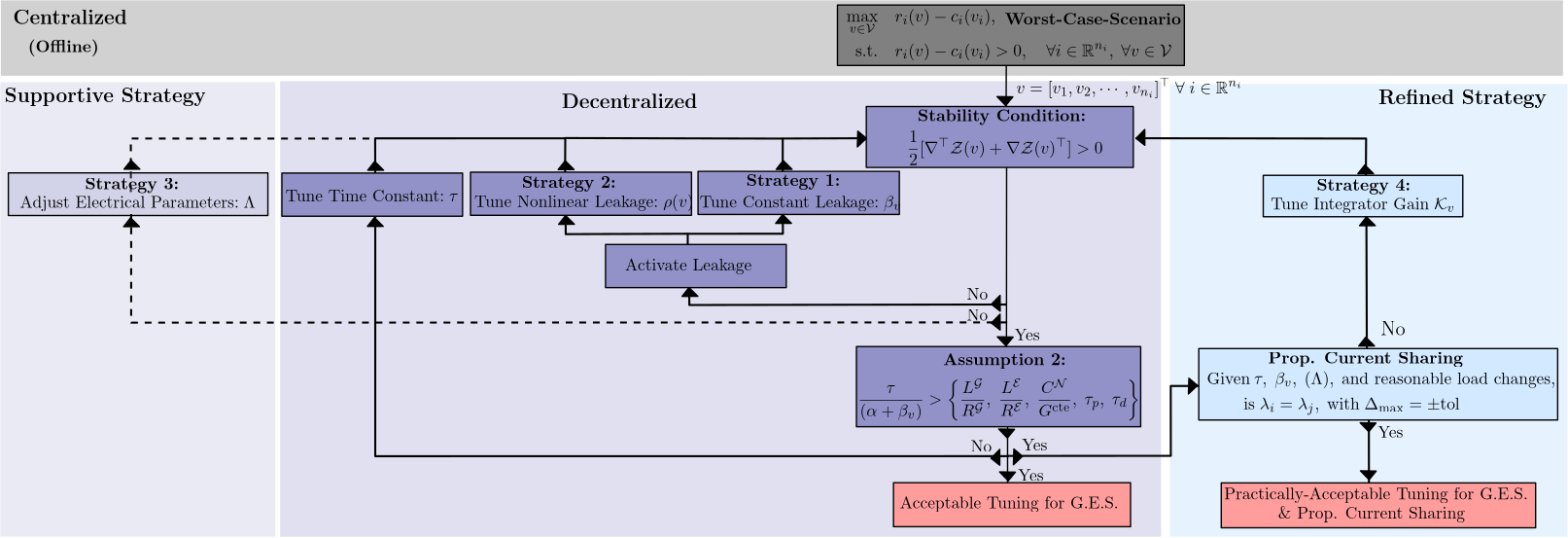}
    \captionsetup{font=small}
    \caption{Tuning Strategy for G.E.S. and Proportional Current Sharing}
    \label{fig:tuning_alg}
\end{figure*}

%% Centralized/decenralized - den er vel begge. - nei, kanskje den er centralized siden vi må gjennom worst case scenario før vi tuner Bv (decentralized) eller IR(cent) eller tau og alpha som er cent? 

\section{Tuning Guidelines for Stability Guarantees}\label{sec_tuning_alg}

The previous section concludes that the system exponentially converges to steady state when we impose sufficient time-scale separation--following \emph{Assumption} \ref{Asump_time_Const}, and when $\mathcal{Z}(v)$ is guaranteed to be a strictly monotonically increasing function--satisfying the stability condition in \eqref{monotonicity_of_M}. Moreover, \emph{Corollary} \ref{cor:Gersh} emphasizes that in order to satisfy the latter condition, we need to ensure that the Geršgorin discs of \eqref{z_matrix} lies entirely in the left-half plane. However, the conditions in \emph{Corollary} \ref{cor:Gersh} might be a be a bit complicated to verify in practice, as both $\mathrm{H}^\lambda(v)$ and $\mathrm{H}^{I^\mathcal{G}}(v)$ depends on all $v \in \mathbb{R}^{n_i}$ leading to multiple operating points and complicating the search for a general and scalable sufficient tuning. Thus, to guarantee G.E.S. of the DC microgrid under all operating conditions, we propose a \emph{worst-case scenario} optimization framework that determines the range of $v$-values in which stability is not guaranteed. Ensuring satisfactory controller tuning within this range will then guarantee stability for all admissible operating points. 
%Thus, by addressing the worst case, a sufficient decentralized tuning can be obtained such that G.E.S. is ensured across all future operating scenarios.

%If a controller tuning ensures stability in the \emph{worst-case}, then this tuning is guaranteed to be sufficient for all admissible operating points. Hence, the objective of the optimization problem is to identify the operating point at which the system lies the furthest beyond its stability bounds. 

Considering \emph{Corollary} \ref{cor:Gersh}, the optimization problem is designed to find the range of $v$-values, that shift the Geršgorin discs in the associated row, into the left-half plane; i.e., when the center value (diagonal element) is less than the absolute sum of the off-diagonal elements (radius) in the $i$th row--mathematically formulated as follows. Let $\varsigma(v)\triangleq \nabla^\top \mathcal{Z}(v)|_\mathrm{sym}$, $c_i(v_i)\triangleq \varsigma(v)_{ii}$, $r_i(v)\triangleq \sum_{j \neq i}|\varsigma(v)|_{ij}$ and $\mathcal{V}=\{v \in \mathbb{R}^{n_i}:v_\mathrm{min}<v<v_\mathrm{max}\}$ for $v=[v_1, v_2, \cdots, v_{n_i}]^\top \; \forall \; i \in \mathbb{R}^{n_i}$
    \begin{align}
    \begin{split}
        \max_{v \in \mathcal{V}} \quad & 
    r_i(v)-c_i(v_i), \\
    \text{s.t.} \quad & r_i(v)-c_i(v_i)>0, \quad \forall i \in \mathbb{R}^{n_i}, \; \forall v \in \mathcal{V}.
    \end{split}
    \end{align}
The optimization problem is solved offline--as illustrated in Fig.~\ref{fig:tuning_alg}--requiring complete system specifications for a given structure. 
\begin{Remark}
    For any change in system topology, the worst-case $v$-values must be recalculated. However, to generalize the results, one could aim to characterize the evolution of this worst-case range for an expanding system. Introducing a safety factor to scale the optimization results may enable obtaining a \emph{worst-case} scenario applicable to future expansion plans. This approach, however, requires further investigation.
\end{Remark}
%Consequently, any change in system topology necessitates recalculation of the \emph{worst-case} $v$-values. Nonetheless, it may be possible to characterize how this \emph{worst-case} range evolves for an expanding system by potentially introducing a safety factor to scale the optimization results for future expansion plans. This approach, however, require further investigation.

%However, once computed, the result remains valid \ul{without requiring recalculation for future system variations}. %Thus, when the optimizer identifies the \emph{unstable} region, we propose \emph{decentralized} tuning strategies-- depicted in Fig.~\ref{fig:tuning_alg} --to carefully obtain a sufficient control tuning, guaranteeing global exponential convergence to a steady state, while ensuring that the resulting equilibrium deviates by no more than $\pm \mathrm{tol}$ from the \emph{optimal} proportional current sharing, further referred to as \emph{practical} proportional current sharing. 

%When examining the dynamics in \eqref{z_matrix}, it is evident that increasing the leakage function $\mathrm{R}(v)$ enhances positive diagonal dominance. Another approach is to reduce the value of $\Lambda$, which simultaneously decreases the off-diagonal elements and strengthens positive diagonal dominance.
\subsection{Decentralized Tuning Strategies for G.E.S.} \label{dec_tuning}

Once the \emph{worst-case} unstable region is established, careful tuning must be applied to satisfy the conditions of \emph{Corollary} \ref{cor:Gersh}. Consequently, we propose three \emph{decentralized} tuning strategies-- illustrated in Fig.~\ref{fig:tuning_alg}. When examining the dynamics in \eqref{z_matrix}, it is evident that increasing the leakage in the diagonal elements, enhance positive diagonal dominance such that center becomes greater than the radius- thus, ensuring positive definite Geršgorin discs. However, Section \ref{sec_SS_opt} emphasizes that any leakage activation diminishes \emph{optimal} proportional current sharing. Therefore, it is of interest to determine a decentralized tuning that ensures compliance with \emph{Corollary} \ref{cor:Gersh} with minimal leakage activation, to ensure that the resulting equilibrium deviates by no more than $\pm \mathrm{tol}$ from the \emph{optimal} proportional current sharing--further referred to as \emph{practical} proportional current sharing.
%
%Accordingly, we propose three tuning strategies, each designed to optimize different aspects of system performance.
%Thus, when the optimizer identifies the \emph{unstable} region, we propose \emph{decentralized} tuning strategies-- depicted in Fig.~\ref{fig:tuning_alg} --to carefully obtain a sufficient control tuning, guaranteeing global exponential convergence to a steady state, while ensuring that the resulting equilibrium deviates by no more than $\pm \mathrm{tol}$ from the \emph{optimal} proportional current sharing, further referred to as \emph{practical} proportional current sharing. 

\textbf{1) A Less Conservative Tuning Strategy}\\
%However, it is important to note that the \emph{non-permanent} leakage function $\rho(v)$ is uniformly defined for all distributed generators.
%For the first tuning strategy, 
We introduce $\mathcal{B}_v$
%$\mathcal{B}_v=\mathrm{diag}(\mathcal{B}^v_i)\in \mathbb{R}^{n_i \times n_i}$
as a \emph{permanent} leakage term, enabling local tunable leakage for the $i$th DG. In accordance with \emph{Corollary} \ref{cor:Gersh}, it suffices to activate only the minimal subset of individual leakages corresponding to the rows of \eqref{z_matrix} for which the Geršgorin Circle Theorem is not satisfied.

\textbf{2) A Generalized Tuning Strategy}\\
In the general case, obtaining the smallest necessary \emph{permanent} leakages of the $i$th DG requires detailed knowledge of the system and often multiple iterations before reaching optimal tuning configurations. Hence, we tune the \emph{non-permanent} leakage $\rho(v)$--previously designed to activate only when the voltages approach their saturation limits--such that it remains \emph{permanently} active, introducing the smallest necessary leakage in the diagonal across all DGs. As this tuning applies to all DGs without requiring local adjustments to the same extent as the previous strategy, the outcome is more generalized and thereby scalable. 
%\ul{Although this outcome is more generalized and thereby scalable}, the resulting tuning tends to be more conservative by enlarging the stability margins beyond what is strictly necessary. 
Furthermore, a more practical approach is to combine the two strategies by activating the \emph{non-permanent} leakage minimally across all DGs, and introduce \emph{permanent} leakage only to those DGs that continue to violate the Geršgorin Circle Theorem. 

\textbf{3) A Supportive Strategy}\\
Diagonal dominance can be further strengthened by increasing the rated currents ($\Lambda = 1/I^\mathrm{rated}$), thereby reducing the magnitude of the off-diagonal elements. 
%\ul{When minimal leakage is introduced in the diagonal of }\eqref{z_matrix}, diagonal dominance can be further enhanced by increasing the rated currents ($\Lambda = 1/I^\mathrm{rated}$), thereby reducing the magnitude of the off-diagonal elements.This approach requires less leakage activation, improving proportional current sharing. However, this method demands modifications to the electrical specifications, making the resulting tuning more application-specific and particularly suited for higher-rated systems.
Although this approach still necessitates some leakage activation, it requires significantly less, thereby improving proportional current sharing. However, this method relies on electrical parameter modifications, making the resulting tuning more application-specific and particularly suited for higher-rated systems.
\begin{Remark}
    To generalize this supportive strategy, a scalar parameter $\nu=1/ \varsigma>0$ can be introduced in \eqref{complete_sys_compact_a}, \eqref{complete_sys_compact_d}, \eqref{complete_sys_compact_e} multiplying the term $\Lambda$ throughout the dynamics. This enables a uniform scaling of the DGs proportional current contribution (in [p.u.]) while improving adherence with the Geršgorin Circle Theorem. Considering the dynamics in \eqref{z_matrix}, increasing the value of $ \varsigma$ emulates a similar behavior to that observed when the controller is applied to a system with higher rated capacities. However, increasing $ \varsigma$, simultaneously reduces the consensus value $\lambda_s$; i.e., the optimal set-point in steady state, thereby enhancing the sensitivity of the proportional current-sharing objective. Consequently, complicating the selection of a tuning that guarantees practical proportional current sharing. 
\end{Remark}

\noindent\textbf{4) A Refined Tuning Strategy}\\ 
%A performance-oriented tuning strategy? 
%An enhanced tuning strategy? 
%refined tuning strategy? Precision? 
For the \emph{practical} proportional current sharing objective, both of the first two strategies cause the DGs to reach a steady state that deviates from \emph{optimal} current sharing to a degree dictated by the introduced leakages. In section \ref{sec_SS_opt}, \emph{optimal} proportional current sharing is defined when the converged steady state reaches $\bar \lambda_i=\Lambda_i\bar I^\mathcal{G}_i$. Hence, increasing the integrator gain $\mathcal{K}_v$ amplifies this condition, bringing the system closer to \emph{practical} proportional current sharing. However, a larger $\mathcal{K}_v$ simultaneously weakens the positive definiteness of \eqref{z_matrix}. Therefore, combining either of the two first initial tuning strategies, while increasing the integrator gain provides an effective means of balancing stability requirements with the objective of ensuring \emph{practical} proportional current sharing in steady state.

\subsection{Preserving the Time-Scale Separation}\label{tuning_time_scale}

%\begin{Remark}
   % Even in saturated operations, the value of $\alpha$ does not influence the time-constant of the decentralized inner-loop controller $\frac{\tau}{\alpha + \mathcal{B}_v}$. This is due to the $\varepsilon$ introduced by the SPT, where $\varepsilon \dot z = g(v,z)$ contains the nonlinear dynamics of the leakage function $\rho(v)v$, and under the stretched time-scale $\mathbf{t}=t/\varepsilon$, this nonlinearity is negligible. 
%\end{Remark}

While the proposed tuning strategies may satisfy the stability condition in \eqref{monotonicity_of_M}, G.E.S. is guaranteed only if sufficient time-scale separation is maintained, in accordance with \emph{Assumption} \ref{Asump_time_Const}. Activating the \emph{permanent} leakage $\mathcal{B}_v$, triggering the \emph{non-permanent} leakage $\rho(v)$--when voltages reach their saturation limit--or modifying the \emph{non-permanent} leakage to remain continuously active, all influence the convergence rate of the decentralized inner-loop controller-- in accordance with \emph{Assumption} \ref{ass:tau}. Consequently, under the aforementioned tuning strategies, the time-scale separation \emph{Assumption} \ref{Asump_time_Const}, must be updated accordingly:
$$\frac{\tau}{\alpha+\beta^\mathrm{max}_v}> \{\tau_p, \tau_d, \frac{L^\mathcal{G}}{R^\mathcal{G}}, \frac{L^\mathcal{E}}{R^\mathcal{E}}, \frac{C^\mathcal{N}}{G^\mathrm{cte}}\}.$$ 

\begin{assumption}\label{ass:tau}
    The proposed evaluation of the time-scales relies on the approximation where the convergence rate of each state can be defined by the ratio of inertia to damping, provided that sufficient damping is introduced in the associated system dynamics.
    \begin{align}\nonumber
        \tau_i=\frac{\mathcal{I}_i}{\mathcal{D}_i}.
    \end{align}
\end{assumption}
Due to the systems nonlinear characteristics, assessing and comparing eigenvalues—and thereby the \emph{true} convergence rates—requires linearization assumptions, and is therefore equilibrium dependent. Thus, when the state under consideration exhibits relatively strong damping, we let \emph{Assumption} \ref{ass:tau} hold. However, when the dynamical state exhibits negligible or non-existing damping--or when a small damping term is introduced solely for mathematical feasibility--the time constant is defined by the parameter multiplying the time-varying state. Although it might be more conservative than eigenvalue analysis, this assumption provides a valid and practically acceptable estimate of the convergence rates, thus ensuring that sufficient time-scale separation is preserved within substantial margins.

Accordingly, to maintain the validity of the attained stable control tuning, we need to increase $\tau$ to satisfy the underlying time-scale separation in various scenarios. From a stability perspective, increasing the value of $\tau$ will not introduce any complications. However, from a performance perspective, it slows down the decentralized integral controller, thereby reducing the steady state convergence rate.
%Note that this also evidences an additional advantage of the \emph{Supportive Tuning Strategy}; when no leakage terms are activated and $\rho(v)$ retains its initial \emph{non-permanent} characteristics, greater flexibility is afforded in selecting the time constant of the decentralized inner-loop controller. 
\begin{Remark}
    In \eqref{complete_sys_compact_f}, $\mathcal{B}_\zeta$ is included for mathematical purposes and has negligible influence on the power system dynamics. Hence, the time constant of $\dot \zeta$ is defined by the parameter $\tau_d$.
\end{Remark}

%and avoids increasing $\tau$ to comply with the time-scale separation assumption. Consequently, the system reaches the \emph{optimal} steady state--when voltage saturation is avoided--while allowing the decentralized inner-loop to operate at a faster rate compared to previously presented tuning strategies.

%%%%%%%%%%%Fordel at man ikke trenger å øke tau så mye - selv ved å bare aktivere rho - hvis REMARK er valid!! kan ha raskere system 

%%Simulations
%1) system med Bv på alle 
%2) system med active rho og Bv (2)
%3) system med active rho og Ir(2)
%4) system med kun inc Ir og ikke active rho - hvis mulig? 
%% kanskje lage tabell med delta_max? 
%% steady state analyse 

%%in the conclusion -- from a small signal perspective, mye er stabilt! men spørsmålet er hvor langt unna large signal vi skal gå eller hvor sonservativt. Inc. i rated hjelper mye mer enn bv, så det er definisjonsspørsmål til brukeren.

\section{Case Studies} \label{sec_Case_Studies}
\begin{figure}[!b]
    \centering
    \includegraphics[width=.9\columnwidth]{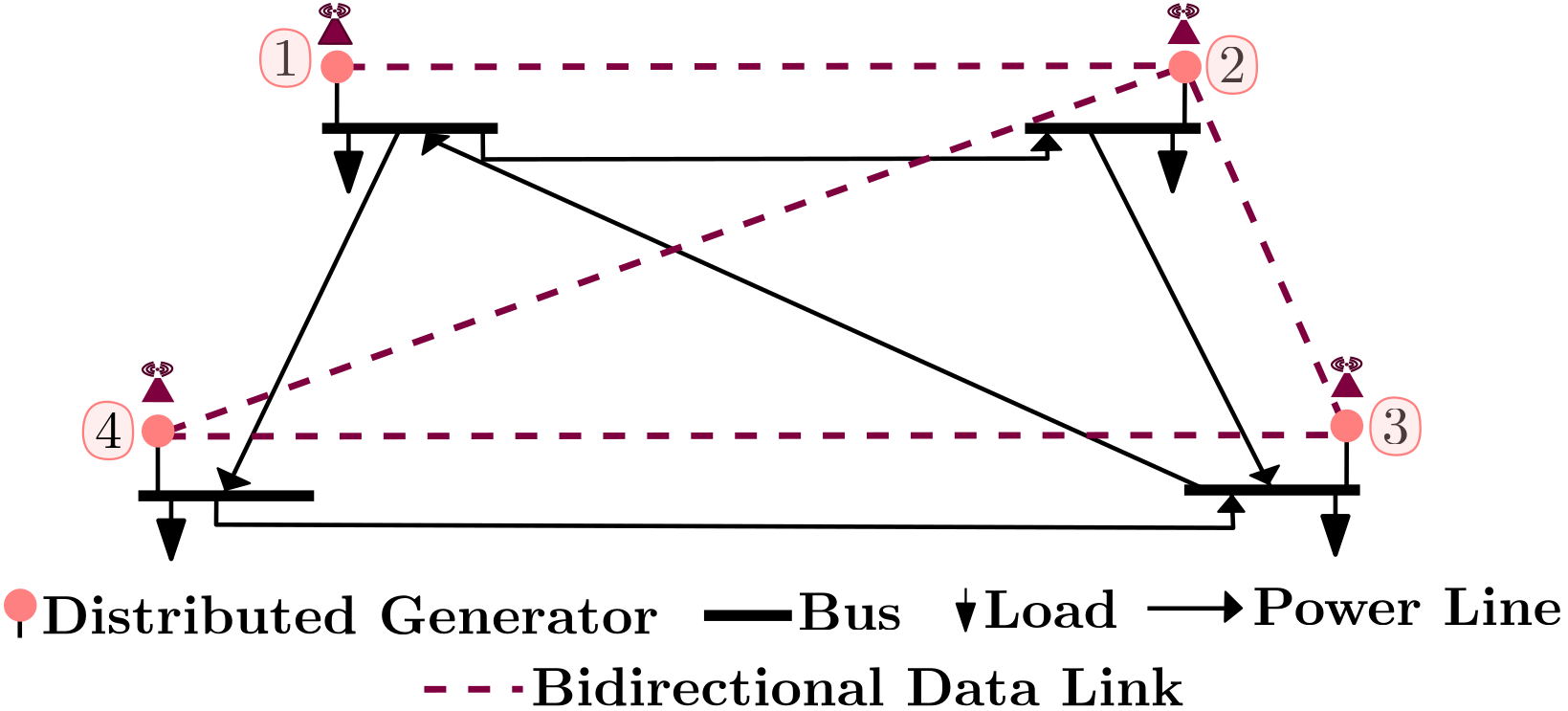}
    \captionsetup{font=small}
    \caption{Case Specific Microgrid}
    \label{Image:Sim_Overview}
\end{figure}
\begin{table}[!b]
    \centering 
    \captionsetup{font=small}
    \caption{Electrical Specifications for Case-Specific MG; resistance ($R$) and inductance ($L$) values are given in per unit (p.u) on a $0.15 \Omega,\; 300 \mu H$ base.} 
    \label{Table_Spec}
    \begin{tabular} {|c|c|c|c|c|}
    \hline
    \multicolumn{5}{|c|}{Generator specifications;  $ i \in \mathcal{G}$}\\
    \Xhline{0.95pt}
    $I_i^\mathrm{rated}$ $[\mathrm{A}]$ & 12 & 4 & 8 & 8\\
    \hline
    $R_i^\mathcal{G} [\mathrm{p.u}]$,$L_i^\mathcal{G} [\mathrm{p.u}]$  & 0.5 & 0.4 & 0.55 & 0.6\\
    \Xhline{0.95pt}
    \multicolumn{5}{|c|}{Load specifications; $k \in \mathcal{N}$}\\
    \Xhline{0.95pt}
    $C_k^\mathcal{N} [\mathrm{F}]$  & \multicolumn{4}{c|}{$2.2 \times 10^{-3}$}\\
    \hline
    $1/G_k^\mathrm{cte} [\Omega]$ & 40 & 30 & 30 & 30 \\
    \hline
    $I_k^\mathrm{cte} [\mathrm{A}]$ & 1 & 1.2 & 0.8 & 1 \\
    \Xhline{0.95pt}
    \multicolumn{5}{|c|}{Power lines specifications; $j \in \mathcal{E}$}\\
    \Xhline{0.95pt}
    \multicolumn{5}{|c|}{
        \begin{tabular}{c|c|c|c|c|c}
        $R_j^\mathcal{E} [\mathrm{p.u}]$, $L_j^\mathcal{E} [\mathrm{p.u}]$ & 1 & 2 & 2 & 1 & 1 \\
        \end{tabular}
        
    } \\
    \hline
    \end{tabular}
\end{table}
%

%After establishing the global exponential stability of the system, we proceed to assess the control performance under optimal operations in steady state according to \emph{Assumption} \ref{Ass_steady state}. \begin{assumption}\label{Ass_steady state} We assume that the distributed controller brings the system to the optimal steady state where the DGs terminal voltage is contained and $\sigma (\lambda)=\Lambda I^\mathcal{G}$ \end{assumption}
The proposed control framework in \eqref{eq:Control-Layer} is tested by means of time-domain simulations in MATLAB/Simulink on a 48-volt DC network admitting the dynamics in \eqref{complete_sys_compact}(a)-(c). The DC microgrid is powered by 4 DGs, interconnected electrically and through communication links according to the interconnection patterns depicted in Fig.~\ref{Image:Sim_Overview}.  The specifications of the generators, loads, and power lines are given in Table \ref{Table_Spec}. For the \emph{base-case} system, the selected control parameters are given below--selected to satisfy \emph{Assumption} \ref{Asump_time_Const}.
\vspace{-.07in}
\begin{equation} \label{init_cont}
\begin{aligned}
\tau &= 5\,\text{s}, \quad      & \tau_p &= 1\,\text{ms}, \quad      & \tau_d &= 10\,\text{ms},\\
k &= 10, \quad & \mathcal{K}_v &= V^*, \quad & \mathcal{B}_v  &= [0,0,0,0], \hspace{0.19
in} \mu=1/100,\\
\alpha&=V_\mathrm{max}, \quad & b &= 5, \quad & \eta &= 1, \hspace{0.5in} \mathcal{B}_\zeta=1e-5.
\end{aligned}
\end{equation}
%%%%%%%%%%%%\mathcal{B}_\zeta=1e-5
The maximum allowed voltage deviation from the nominal voltage, $V_n=48V$, is $5\%$; $V_{\mathrm{max}}=1.05$[p.u]. Note that we choose $\mu=1/100$ to neglect the influence of linear dynamics $\Lambda^\top \mu \lambda$, included in the control actuator \eqref{complete_sys_compact_a} for mathematical purposes. Moreover, for \emph{practical} proportional current sharing, we permit a practically acceptable $\pm 5\%$ deviation from the \emph{optimal} steady state operating point. Throughout the subsequent case studies, we aim to find adequate tuning of our controller, discussing various acceptable tuning values of $\mathcal{B}_v$, $\rho(v)$, $\mathcal{K}_v$ (and $I^\mathrm{rated}$) for scalable G.E.S..

%\emph{Corollary} \ref{Coroll_alpha} highlights that achieving global exponential stability guarantees for large-signal disturbances is highly dependent on reducing the inter-dependencies in the electrical system such that $\mathcal{M}(\lambda)$ is strictly monotonically increasing. Accordingly, through the subsequent case studies, we aim to find adequate tuning of our controller -- tested for various electrical system parameters -- to guarantee G.E.S. following the stability conditions in \emph{Theorem} \ref{theorem_GES} while striving to satisfy optimal steady-state operation of the DC microgrid, where both voltage containment and proportional current sharing are maintained, as established in \emph{Lemma} \ref{lemma_Optimal_steady_state}.

\subsection{Case Study 1: Sufficient Tuning to Satisfy Corollary~\ref{cor:Gersh}} \label{sec:CS1}
\begin{figure*}[t!]
    \centering
    \includegraphics[width=\textwidth]{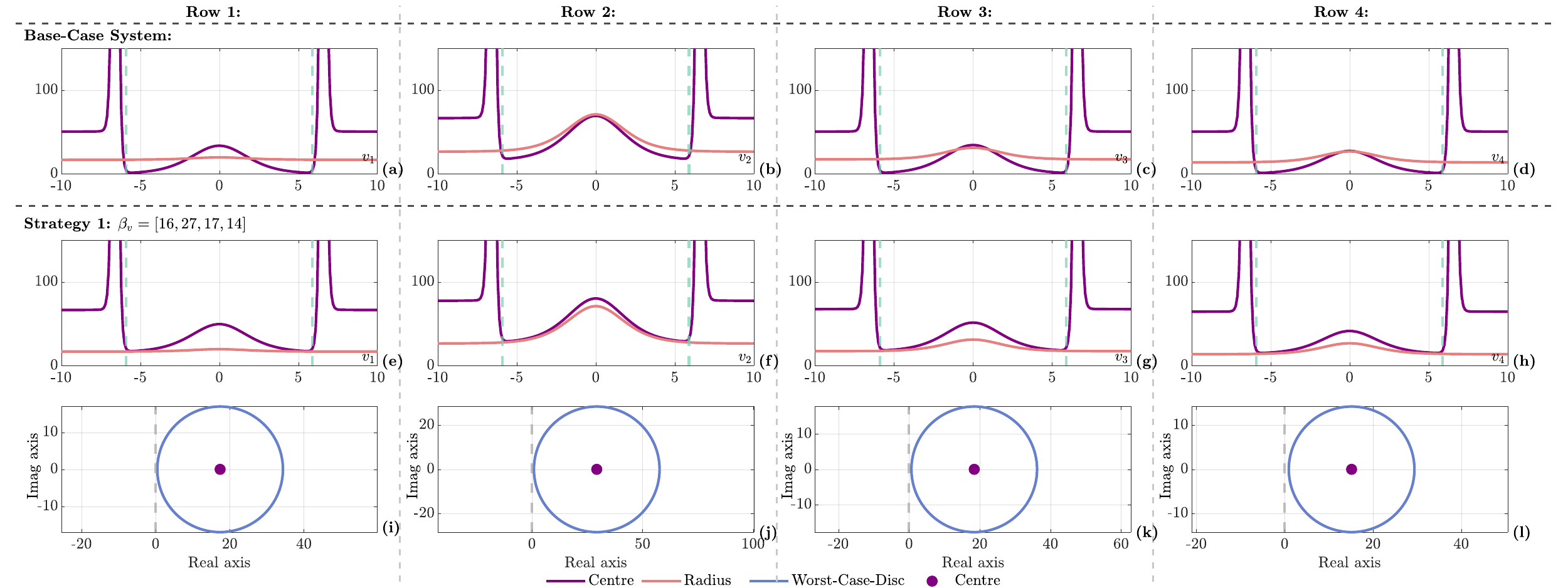}
    \captionsetup{font=small}
    \caption{Gashgorin Circle Theorem Plots: Base-case system (first row), tuned system with individual permanent leakage ($\mathcal{B}_v$) (second and third row); (a)-(h) center v.s. radius for each row in $\mathcal{Z}(v)$; (i)-(l) Geršgorin disc of each row in $\mathcal{Z}(v)$ for the worst case scenario}
    \label{Fig:Gash_X12}
\end{figure*}

Before testing the proposed control design on our case-specific DC MG, we need to obtain sufficient tuning configurations that guarantee G.E.S.. First, we solve the optimization problem to find the set of unstable $v$-values. Subsequently, following the strategies in Fig.~\ref{fig:tuning_alg}, we propose three sufficient controller tunings to ensure adherence with \emph{Corollary} \ref{cor:Gersh}. Hereafter, the \emph{base-case} system refers to initial control configurations--parameter values given in \eqref{init_cont}--preliminary implemented without careful tuning, and consequently, without any formal stability guarantees. When solving the optimization problem for this \emph{base-case} system, the optimizer converges to the \emph{worst-case} scenario when the decentralized integral controller states are $v_{ii}=[5.54, 5.42, 5.51, 5.47]^\top$ and $v_{i,l}=0$ in the corresponding $i$th row. This behavior is further illustrated in Fig.~\ref{Fig:Gash_X12}(a)–(d), where the center (diagonal element) and radius (absolute sum of the off-diagonal elements) are plotted for varying $v_{i}$-values on the $i$th row in \eqref{z_matrix}. It is evident that the $v_{i}$-value associated with the most unstable case, i.e., when the plotted center exhibits its maximum negative deviation from the radius, corresponds with the solution of the \emph{worst-case} optimizer. Furthermore, the green dashed lines indicate the activation of the \emph{non-permanent} leakage, highlighting the need for sufficient leakage to maintain stability as the center then becomes more positive than the radius.
Following Section \ref{dec_tuning}, we propose three tuning strategies to guarantee that the \emph{worst-case} scenario satisfies \emph{Corollary} \ref{cor:Gersh}:
    \begin{IEEEeqnarray}{lCl}
    \begin{aligned} &\textbf{Strategy 1:}\hspace{.1in}\rho(v),\hspace{.04in} \mathcal{B}_v=[16,46,17,14],\hspace{.04in}I^\mathrm{rated}=[12, 4, 8, 8],\hspace{.04in}\mathcal{K}_v=V^*\\
    &\textbf{Strategy 2:}\hspace{.04in}\rho^*(v),\hspace{.04in}\mathcal{B}_v=[0,16,0,0], \hspace{.24in}I^\mathrm{rated}=[12, 4, 8, 8],\hspace{.04in}\mathcal{K}_v=V^*\\ &\textbf{Strategy 3:}\hspace{.04in}\rho^*(v),\hspace{.04in}\mathcal{B}_v=[0,0,0,0],\hspace{.3in}I^\mathrm{rated}=[12, 8, 8, 8], \hspace{.04in}\mathcal{K}_v=V^*
    \end{aligned}\nonumber
    \end{IEEEeqnarray}
In the first tuning strategy, we activate the \emph{permanent} leakage $\mathcal{B}_v$ locally adjusted for all DGs. The corresponding plots in Fig.~\ref{Fig:Gash_X12} (e)-(h) confirm that this represents the minimal individual tuning required to satisfy the condition: $\text{center}>|\text{radius}|$, in each row. Moreover, the resulting Geršgorin discs all lie in the right-half plane, as depicted in Fig.~\ref{Fig:Gash_X12} (i)-(l). Hence, under this tuning, \emph{Corrolary} \ref{cor:Gersh} is guaranteed for the \emph{worst-case}.
 %%%hovwever not concluding upon G.E.S. following the stability condition in \eqref{monton} yet, as we need to tune $\tau$ before we can conclude -- and maybe check the \emph{practical} proportional current sharing. 

\begin{figure}[!t]
    \centering
    \includegraphics[width=.78\columnwidth]{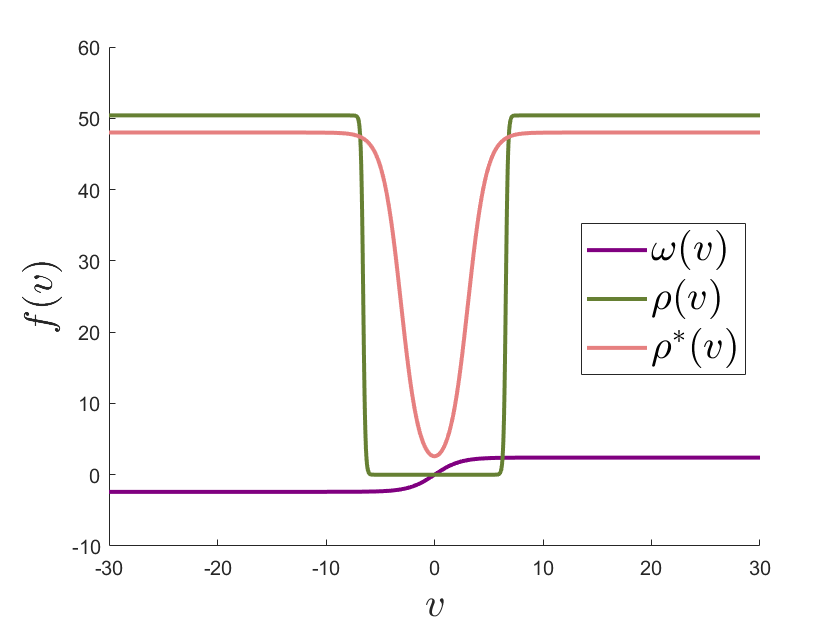}
    \captionsetup{font=small}
    \caption{Nonlinear Leakage Functions}
    \label{Fig:lekages}
\end{figure}

In accordance with the second tuning strategy, we adjust the \emph{non-permanent} leakage function, $\rho(v)$, to remain continuously active-—thereby ensuring uniform leakage activation across all DGs, while aiming to keep the magnitude small enough not to significantly hinder proportional current sharing. For proper voltage containment, the slope of the saturation curve is adjusted while preserving the feature that $\rho(v)$ attains its minimum when voltages are near the midpoint of their allowable range. Second, the leakage gain $\alpha$, is carefully reduced to satisfy the time-scale assumption, while remaining sufficiently active to facilitate achieving the conditions of Corollary \ref{cor:Gersh}. Fig.~\ref{Fig:lekages} illustrates the correlation between the nonlinear voltage controller $\omega(v)$, the initial \emph{non-permanent} leakage $\rho(v)$, and the modified continuously active leakage $\rho^*(v)$ when taking the following values
\vspace{-.05in}
\[
\begin{alignedat}{3}
\alpha&=48    &\quad b &= 0.58 &\quad \eta &= 0.47.\\
\end{alignedat}
\]
\vspace{-.2in}

\begin{figure*}[b!]
    \centering
    \includegraphics[width=\textwidth]{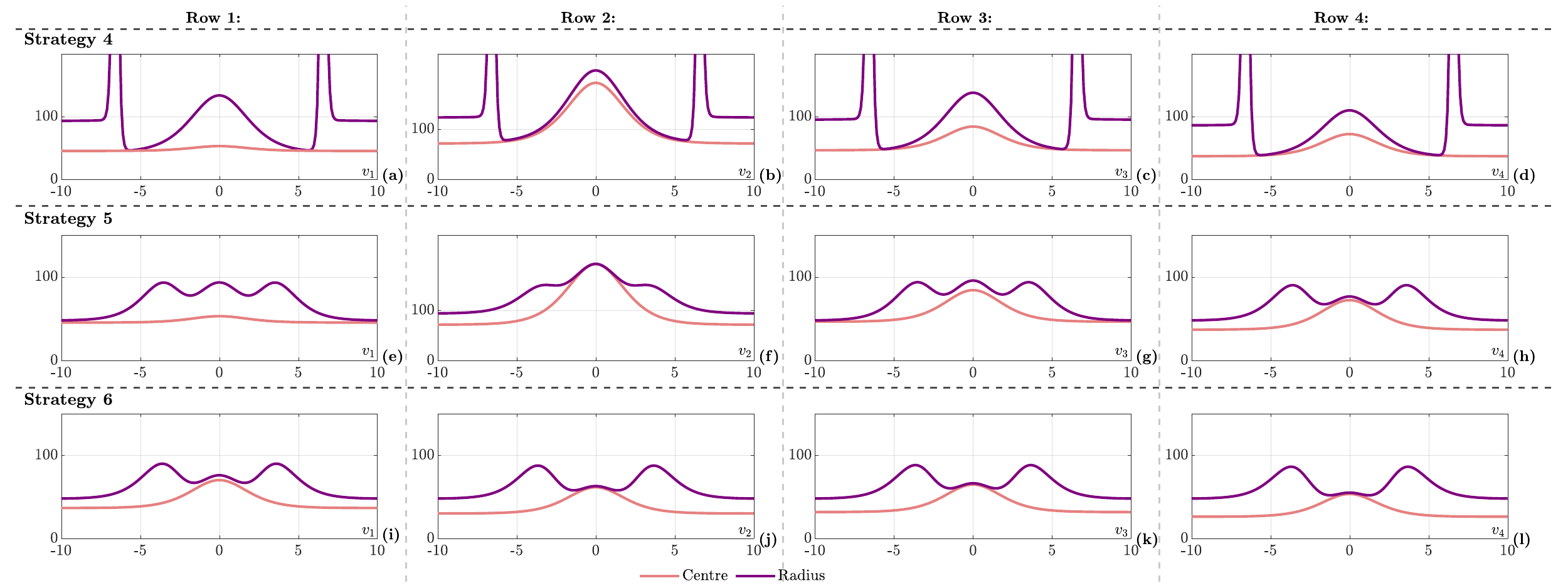}
    \captionsetup{font=small}
    \caption{Gashgorin Circle Theorem Plots: Tuned system for \emph{practical} proportional current sharing when $\mathcal{K}_v=130$; (a)-(d) Strategy 4; (e)-(h) Strategy 5; (i)-(l) Strategy 6}
    \label{Fig:Gash_X8}
\end{figure*}

Fig.~\ref{Fig:Gash_X8}(a)-(d) depicts that the first, third, and fourth rows in \eqref{z_matrix} satisfy the conditions of the Geršgorin Circle Theorem under this tuning; however, the condition is not met for the second row. To address this, we propose two options: either activate the \emph{permanent} leakage for the second generator $\mathcal{B}_2^v=16$ (Strategy 2), or increase its rated current by $4[A]$; $I_2^\mathrm{rated}=8$ (Strategy 3). Fig.~\ref{Fig:Gash_X8}(e)-(f) illustrates that when $\mathcal{B}_2^v$ is activated, the center becomes greater than the radius for all $v_{2}$-values, and the \emph{worst-case} Gashgorin disc lies in the right-half plane--indicating sufficient tuning configurations. Following the second option (Strategy 3), the results depicted in Fig.~\ref{Fig:Gash_X8}(g)–(h) exhibit a similar response, confirming that the Geršgorin Circle Theorem conditions are met for all rows. 

Although this tuning (both Strategy 2 and 3) offers a more generalized and scalable solution--requiring less generator-specific adjustments--it is more conservative. As indicated in Fig.~\ref{Fig:Gash_X8}(a), $\rho^*(v_1)$ is activated to an extent exceeding what is necessary to satisfy the Geršgorin Circle Theorem. %Moreover, this may lead to larger deviations from optimal proportional current sharing.
\subsection{Case Study 2: System Response and Practical Proportional Current Sharing} \label{sec:CS2}

\begin{figure*}[t!]
    \centering
    \includegraphics[width=\textwidth]{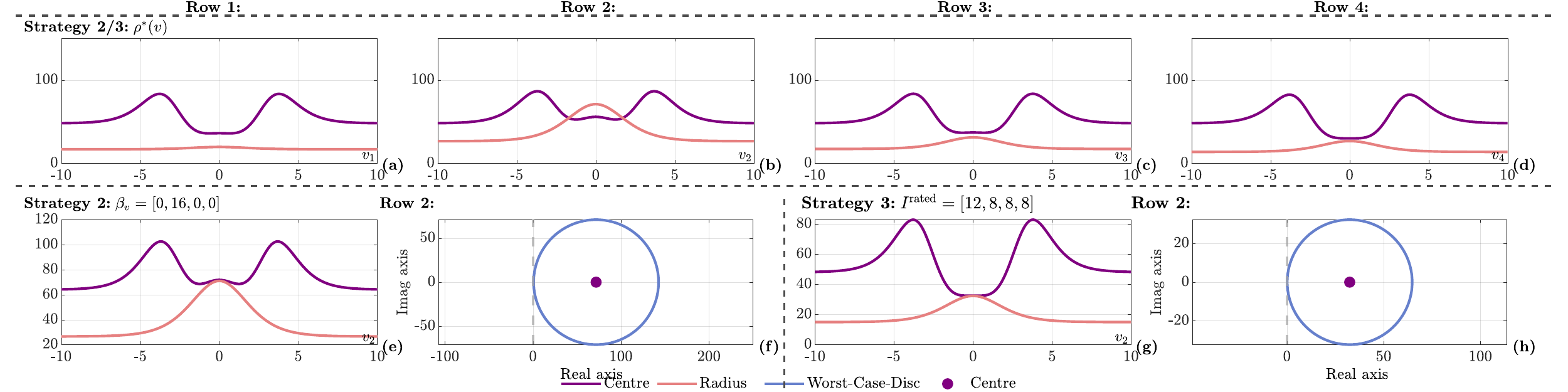}
    \captionsetup{font=small}
    \caption{Gashgorin Circle Theorem Plots: Tuned system with modified non-permanent leakage ($\rho(v)$) with; additional permanent leakage ($\mathcal{B}_v$) (second row, column 1\&2),  increased rated currents (second row, column 3\&4); (a)-(e), (g) center v.s. radius for relevant rows in $\mathcal{Z}(v)$; (f),(h) Geršgorin disc of relevant row in $\mathcal{Z}(v)$ for the worst case scenario}
    \label{Fig:Opt_Gash_X12}
\end{figure*}
%% presentere kjapr hvilke events, hvor mye vi øker og hvorfor

This case study evaluates the performance and effectiveness of the proposed control framework in \eqref{complete_sys_compact} when applied to the case-specific DC microgrid under the proposed tuning strategies. When $t \in [12, 89]$, we impose various load variations, increasing and decreasing $I_i^\mathrm{cte}$ and $G^\mathrm{cte}_i$ up to $50\%$. When $t \in [95, 170]$, we disconnect and reconnect DG1, load 2, and the power line between load 3 and 4. The following simulation studies--depicted in Fig.~\ref{Fig:Sim}--are conducted to evaluate the controllers ability to ensure voltage containment and proportional current sharing under the proposed tuning strategies, and under various operating conditions, including both small- and large-signal disturbances. Moreover, we aim to assess the controllers robustness and validate its scalability through the different plug-and-play scenarios. 

Although the proposed tuning strategies (1, 2, and 3) satisfy the Geršgorin Circle Theorem and thereby fulfill the stability condition in \eqref{monotonicity_of_M}, it is essential to verify the time-scale separation--to guarantee G.E.S.--before conducting the simulations. In accordance with Section \ref{tuning_time_scale}, the time constant of the decentralized controller is adjusted as detailed below. Note that the table includes the time constants for Strategies 4, 5, and 6, as well as illustrating the system responses under these tuning strategies in Fig.~\ref{Fig:Sim}(c), (f), and (i)--which will be further discussed in the subsequent case study. 
\begin{table} [h!]
    \centering 
    \captionsetup{font=small}
    \caption{$\tau$ values to satisfy the time-scale separation assumption} \label{Table_equilibrium}
    \resizebox{0.95\columnwidth}{!}{
    \begin{tabular} {|c|c|c|c|c|c|c|} 
    \hline
     & Strategy 1 & Strategy 2 & Strategy 3 & Strategy 4 & Strategy 5 & Strategy 6 \\
    \hline
    $\tau[s]$ & 9 & 6 & 5 & 11 & 9 & 5\\
    \hline
    \end{tabular}
    }\label{table_tau}
\end{table}

Fig.~\ref{Fig:Sim}(a), (c), and (e), illustrate that the controller ensures voltage containment across all imposed events--within substantial safety margins. However, when considering the second control objective--\emph{practical} proportional current sharing--the response of Strategy 1, illustrated in Fig.~\ref{Fig:Sim}(b), shows that the system converges to a steady state that deviates significantly beyond the practically acceptable tolerance ($\pm 5\%$), due to the necessarily high individual leakage values required to guarantee G.E.S.. For \emph{optimal} proportional current sharing, the ideal system should exhibit a response where the DGs converge to a steady state in which their integration errors approach zero. However, leakage activation prevents complete convergence in steady state, resulting in a percentage deviation from the optimal current sharing condition, given by: $\Delta_\mathrm{max}= \left[\frac{(\lambda_i-\Lambda_i I_i^\mathcal{G})}{\lambda_i}\right]_\mathrm{t}$, measured for each time-step $t$. Fig.~\ref{Fig:Sim}(d) demonstrates that the second tuning strategy provides a considerably improved current-sharing performance; however, not within an acceptable tolerance in all imposed system events. Among the tested configurations, Strategy 3 exhibits the best overall current sharing performance--depicted in Fig.~\ref{Fig:Sim} (h)-- as anticipated, since its tuning approach does not rely on activating any local leakage terms to guarantee stability. Nevertheless, this strategy requires the controller to be implemented on a system with higher generator capacities to maintain its performance advantages.

%%Mp si hyvordan deviation er regnet ut, og hvordan optimal skal se ut!
%% If the focus is on scalability and prop.power sharing--choose option 2!!

%% Define that \emph{optimal} prop current sharing is when the integrator error stabilizes around zero for all DGs

\subsection{Case Study 3: Refined tuning for Practical Proportional Current Sharing} \label{sec:CS3}

For \emph{practical} proportional current sharing, we aim to identify a control tuning that minimizes the percentage deviation from \emph{optimal} proportional current sharing. Thus, we propose three new tuning configurations by combining the \emph{refined} tuning strategy in Section \ref{sec_tuning_alg} with the three previously proposed Strategies 1, 2 and 3. Specifically, we increase the integrator gain $\mathcal{K}_v$, after which sufficient retuning is required to satisfy the Geršgorin Circle Theorem. Hence, we modify Strategies 1, 2, and 3 when $\mathcal{K}_v=130$, which results in the proposed Strategies 4, 5, and 6, respectively--each defined by the following tuning configurations.
\begin{IEEEeqnarray}{lCl}
    \begin{aligned} &\textbf{Strategy 4:}\hspace{.1in}\rho(v),\hspace{.04in} \mathcal{B}_v=[43,73,45,36],\hspace{.04in}I^\mathrm{rated}=[12, 4, 8, 8],\hspace{.04in}\mathcal{K}_v=130\\
    &\textbf{Strategy 5:}\hspace{.04in}\rho^*(v),\hspace{.04in}\mathcal{B}_v=[0,46,0,0], \hspace{.24in}I^\mathrm{rated}=[12, 4, 8, 8],\hspace{.04in}\mathcal{K}_v=130\\ &\textbf{Strategy 6:}\hspace{.04in}\rho^*(v),\hspace{.04in}\mathcal{B}_v=[0,0,0,0],\hspace{.1in}I^\mathrm{rated}=[12, 11, 11, 11], \hspace{.04in}\mathcal{K}_v=130
    \end{aligned}\nonumber
    \end{IEEEeqnarray}

 \begin{figure*}[b]
    \centering
    \includegraphics[width=\textwidth]{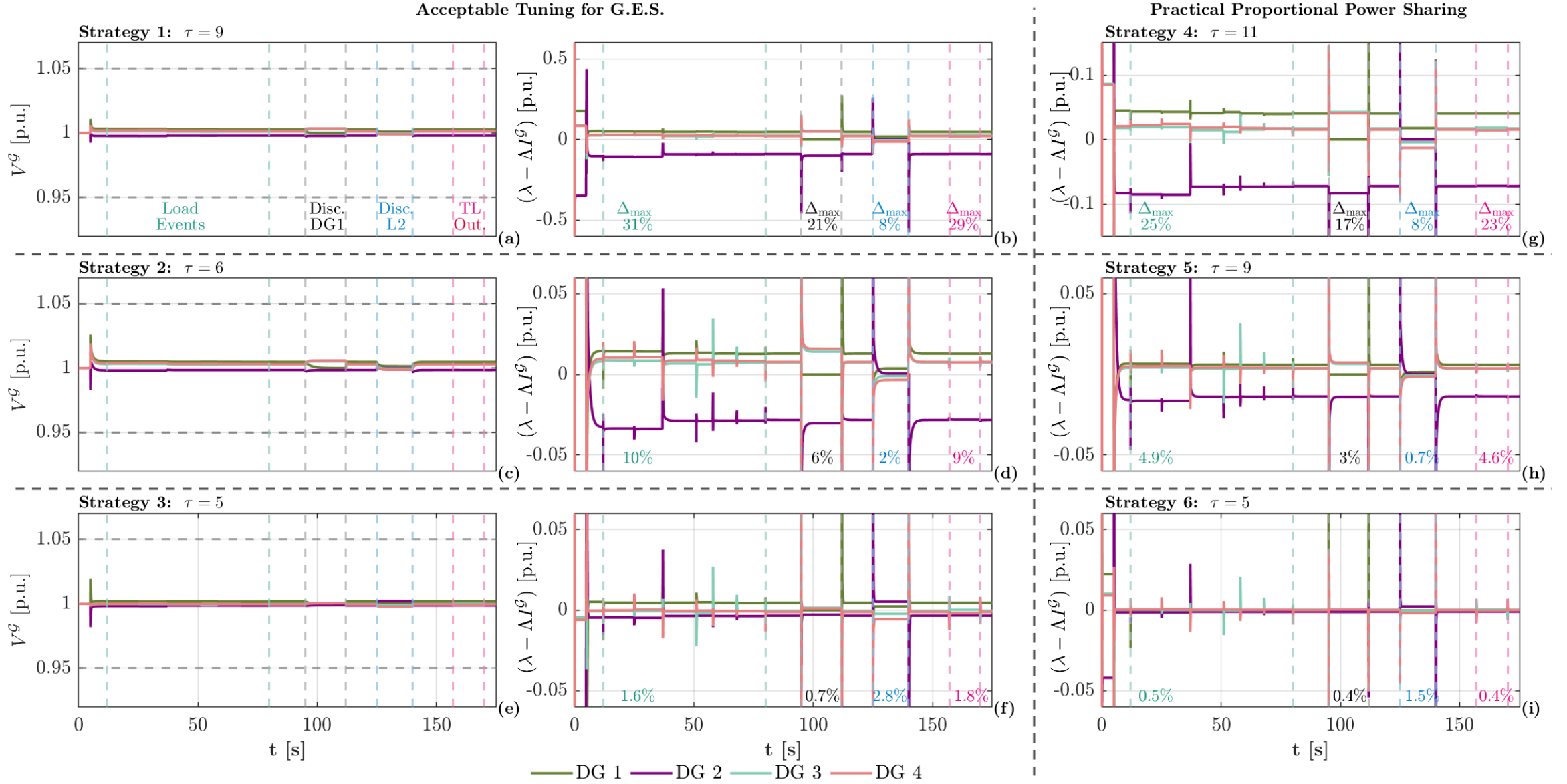}
    \captionsetup{font=small}
    \caption{Simulation results under tuning Strategy 1 (a)-(b), tuning Strategy 2 (c)-(d), tuning Strategy 3 (e)-(f), tuning Strategy 4 (g), tuning Strategy 5 (h), tuning Strategy 6 (i); (a), (c), (e) generators voltages; (b), (d), (f), (g), (h), (i) integrations errors; Imposed load changes: $t=5:$  activation of distributed controller, $t=12: \mathrm{Inc.} I_1^\mathrm{cte} 50\%$, $t=25: \mathrm{Red.} I_4^\mathrm{cte} 13\%$, $t=37: \mathrm{Red.} G_1^\mathrm{cte} 40\%$, and $\mathrm{Red.} G_2^\mathrm{cte} 22\%$, $t=51: \mathrm{Inc.} I_3^\mathrm{cte} 17\%$, $t=58: \mathrm{Red.} I_3^\mathrm{cte} 45\%$, $t=68: \mathrm{Red.} G_3^\mathrm{cte} 7\%$, $t=80: \mathrm{Inc.} G_4^\mathrm{cte} 10\%$,; Imposed disconnections: $t\in[95,112]$ Disc. DG1, $t\in[125,140]$ Disc. load 2, $t\in[157, 170]$ Disc. TL between load 3 and 4 }
    \label{Fig:Sim}
\end{figure*}

As previously discussed, increasing the integrator gain $\mathcal{K}_v$ improves \emph{practical} proportional current sharing. However, it simultaneously weakens the overall stability condition. This effect is further confirmed through the Geršgorin circle theorem plots of the \emph{base-case} system when $\mathcal{K}_v=130$ (omitted here for brevity), which causes $\text{center}<|\text{radius}|$ for all corresponding rows around the \emph{worst-case} $v_i$-values.
%and for the second DG, the Geršgorin circle theorem is never satisfied under this tuning. 
Consequently, Fig.~\ref{Fig:Opt_Gash_X12} demonstrates that under the proposed tuning Strategies 4, 5, and 6, the Geršgorin circle conditions are satisfied across all associated rows. Specifically, Strategy 4 activates the minimum required individual \emph{permanent} leakage, while Strategies 5 and 6 involve employing the modified $\rho^*(v)$ function and adjusting the \emph{permanent} leakage of the second DG (Strategy 5), or increasing the rated currents of the second, third, and fourth DGs (Strategy 6). It is noteworthy that when the integrator gain is increased, enhancing the rated capacity of DG2 alone—as in Strategy 2—is no longer sufficient. In this case, the controller must be implemented on a system where all DGs exhibit approximately equal and adequately high-rated capacities to ensure stability.

Subsequently, Fig.~\ref{Fig:Sim}(g)–(i) illustrates the system response under the proposed \emph{refined} tuning strategies. To ensure sufficient time-scale separation, the time constant of the decentralized controller $\tau$ is increased in accordance with Assumption \ref{ass:tau}, with the corresponding values provided in Table \ref{table_tau}. When considering \emph{practical} proportional current sharing, a similar conclusion can be drawn as in the previous case study. When applying Strategy 4, the converged steady state deviates significantly from optimal operation, whereas Strategy 5 reduces this deviation, and Strategy 6 provides the best overall performance. However, when comparing the results with the previous case study, all \emph{refined} tuning configurations demonstrate improved current-sharing performance. Furthermore, Strategy 5 now guarantees convergence to a steady state with a $\pm 5\%$ deviation from \emph{optimal} operations, during all imposed system events. Hence, it can be concluded that Strategy 5 provides acceptable and sufficient tuning conditions without relying on any electrical specifications, as necessary in Strategies 3 and 6.

\section{Conclusion and Future Work}
\label{conc}

In conclusion, this paper proposes a nonlinear nested distributed control framework for DC microgrids, designed to ensure proportional current sharing and voltage containment in steady state. By employing singular perturbation theory and Lyapunov theory, we establish a \emph{scalable} global exponential stability (G.E.S.) certificate under some stability constraints and sufficient time-scale separation between the decentralized inner-loop and the rest of the system. Specifically, the decentralized controller is required to operate at a slower time-scale. To satisfy the stability constraints, we first propose an optimization algorithm that identifies the \emph{worst-case} scenario and subsequently propose a tuning scheme to guarantee G.E.S. for this case--thus ensuring stability for all admissible operating conditions. The proposed tuning framework encompasses several strategies depending on practical engineering objectives, enabling either a less conservative design, a more general and scalable implementation, a focus on optimizing the proportional current sharing, or enhanced flexibility with respect to electrical specifications.

Furthermore, we validate the proposed control framework on a case-specific DC microgrid and introduce three tuning strategies to guarantee G.E.S under the identified \emph{worst-case} scenario. Focusing on proportional current sharing, we further refine these strategies--yielding three additional tuning configurations that bring the system performance closer to \emph{practical} proportional current sharing, i.e., allowing a permissible deviation from the \emph{optimal} operating point within a tolerance of $\pm$tol. 

Compared to the previously established G.E.S. certificates in \cite{Poppi_J1} and \cite{Poppi_J2}, the results of this study preserve less conservative stability margins by relaxing the requirement that the outer-loop controller and its communication must be either faster or slower than the electrical system and its decentralized integral controller. This result allows for either very fast communication, or operational rates within the same time-scales as the electrical system, or even slower, thus avoiding practical limitations in the previous attempts. The stability certificate is therefore less application specific --compared to the result in \cite{Poppi_J1}-- and does not rely on any parameter specification as in \cite{Poppi_J2}. However, stability is still preserved with robust margins, characterized by the time-scale separation approximation, resulting in more conservative convergence rates. 

Moreover, depending on the applied tuning configurations, stability may be preserved beyond whats strictly necessary in order to guarantee G.E.S.. For further studies, refining the \emph{worst-case} optimization problem to make it less system-dependent and valid for expandable systems could improve its general applicability for future applications. Moreover, containing the linear term introduced in the primary controller--introduced solely for mathematical structures--could result in a more precise voltage containment configuration; however, this would require reconfiguring the Lyapunov function and revising the stability proof.

\balance
\bibliographystyle{IEEEtran.bst}
\bibliography{IEEEabrv,Refs}

\end{document}